\documentstyle[12pt,epsf]{article} \hoffset -0.5in \textwidth 6.00in \textheight
8.5in  \parskip 7pt \openup1\jot \parindent=0.5in
\newskip\humongous \humongous=0pt plus 1000pt minus 1000pt
\def\caja{\mathsurround=0pt}
\def\eqalign#1{\,\vcenter{\openup1\jot \caja
        \ialign{\strut \hfil$\displaystyle{##}$&$
        \displaystyle{{}##}$\hfil\crcr#1\crcr}}\,}
\newif\ifdtup

\def\eqright #1\cr{\noalign{\hfill$\displaystyle{{}#1}$}}
\def\eqleft #1\cr{\noalign{\noindent$\displaystyle{{}#1}$\hfill}}
\def\oldreffmt#1{\rlap{[#1]} \hbox to 2\parindent{}}

\def\figfmt#1{\rlap{Figure {#1}} \hbox to 1in{}}

\def\VEV#1{\left\langle #1\right\rangle}

\def\sectioneq{\def\theequation{\thesection.\arabic{equation}}{\let
\holdsection=\section\def\section{\setcounter{equation}{0}\holdsection}}}%

\newcounter{holdequation}

\def\auto{\eqno(\refstepcounter{equation}\theequation)}
\def\begineq #1\endeq{$$ \refstepcounter{equation}\eqalign{#1}\eqno
	(\theequation) $$}
\def\contlimit{\,{\hbox{$\longrightarrow$}\kern-1.8em\lower1ex
\hbox{${\scriptstyle (a\rightarrow0)}$}}\,}
\def\centeron#1#2{{\setbox0=\hbox{#1}\setbox1=\hbox{#2}\ifdim
\wd1>\wd0\kern.5\wd1\kern-.5\wd0\fi
\copy0\kern-.5\wd0\kern-.5\wd1\copy1\ifdim\wd0>\wd1
\kern.5\wd0\kern-.5\wd1\fi}}
\def\centerover#1#2{\centeron{#1}{\setbox0=\hbox{#1}\setbox
1=\hbox{#2}\raise\ht0\hbox{\raise\dp1\hbox{\copy1}}}}
\def\centerunder#1#2{\centeron{#1}{\setbox0=\hbox{#1}\setbox
1=\hbox{#2}\lower\dp0\hbox{\lower\ht1\hbox{\copy1}}}}
\def\lsim{\;\centeron{\raise.35ex\hbox{$<$}}{\lower.65ex\hbox
{$\sim$}}\;}
\def\gsim{\;\centeron{\raise.35ex\hbox{$>$}}{\lower.65ex\hbox
{$\sim$}}\;}
\def\st#1{\centeron{$#1$}{$/$}}
\def\super#1{\ifmmode \hbox{\textsuper{#1}}\else\textsuper{#1}\fi}
\def\textsuper#1{\newcount\holdspacefactor\holdspacefactor=\spacefactor
$^{#1}$\spacefactor=\holdspacefactor}
\def\getcite#1,{\advance\citenumber by1
\def\getcitearg{#1}\def\lastarg{@}
\ifnum\citenumber=1
\ref{#1}\let\next=\getcite\else\ifx\getcitearg\lastarg\let\next=\relax
\else ,\ref{#1}\let\next=\getcite\fi\fi\next}
\def\pom{{\rm P\kern -0.53em\llap I\,}}
\def\spom{{\rm P\kern -0.36em\llap \small I\,}}
\def\sspom{{\rm P\kern -0.33em\llap \footnotesize I\,}}

\relax

\def\auto{\eqno(\refstepcounter{equation}\theequation)}
\def\begineq #1\endeq{$$ \refstepcounter{equation}\eqalign{#1}\eqno
	(\theequation) $$}
\def\contlimit{\,{\hbox{$\longrightarrow$}\kern-1.8em\lower1ex
\hbox{${\scriptstyle (a\rightarrow0)}$}}\,}
\def\centeron#1#2{{\setbox0=\hbox{#1}\setbox1=\hbox{#2}\ifdim
\wd1>\wd0\kern.5\wd1\kern-.5\wd0\fi
\copy0\kern-.5\wd0\kern-.5\wd1\copy1\ifdim\wd0>\wd1
\kern.5\wd0\kern-.5\wd1\fi}}
\def\centerover#1#2{\centeron{#1}{\setbox0=\hbox{#1}\setbox
1=\hbox{#2}\raise\ht0\hbox{\raise\dp1\hbox{\copy1}}}}
\def\centerunder#1#2{\centeron{#1}{\setbox0=\hbox{#1}\setbox
1=\hbox{#2}\lower\dp0\hbox{\lower\ht1\hbox{\copy1}}}}
\def\lsim{\;\centeron{\raise.35ex\hbox{$<$}}{\lower.65ex\hbox
{$\sim$}}\;}
\def\gsim{\;\centeron{\raise.35ex\hbox{$>$}}{\lower.65ex\hbox
{$\sim$}}\;}
\def\st#1{\centeron{$#1$}{$/$}}
\def\super#1{\ifmmode \hbox{\textsuper{#1}}\else\textsuper{#1}\fi}
\def\textsuper#1{\newcount\holdspacefactor\holdspacefactor=\spacefactor
$^{#1}$\spacefactor=\holdspacefactor}
\def\getcite#1,{\advance\citenumber by1
\ifnum\citenumber=1
\ref{#1}\let\next=\getcite\else\ifx#1@\let\next=\relax
\else ,\ref{#1}\let\next=\getcite\fi\fi\next}
\def\upon #1/#2 {{\textstyle{#1\over #2}}}
\relax

\def\mainhead#1{\setcounter{equation}{0}\addtocounter{section}{1}
  \vbox{\begin{center}\large\bf #1\end{center}}\nobreak\par}
\sectioneq

\def\til#1{\centeron{\hbox{$#1$}}{\lower 2ex\hbox{$\char'176$}}}
\def\tild#1{\centeron{\hbox{$\,#1$}}{\lower 2.5ex\hbox{$\char'176$}}}
\def\sumtil{\centeron{\hbox{$\displaystyle\sum$}}{\lower
-1.5ex\hbox{$\widetilde{\phantom{xx}}$}}}

\def\pom{{\rm P\kern -0.53em\llap I\,}}
\def\spom{{ I\kern -0.20em\llap \small P\,}}
\def\sspom{{\rm P\kern -0.33em\llap \footnotesize I\,}}

\newcommand{\bit}{\begin{itemize}}
\newcommand{\eit}{\end{itemize}}

\newcommand{\beq}{\begin{equation}}
\newcommand{\eeq}{\end{equation}}
\newcommand{\beqa}{\begin{eqnarray}}
\newcommand{\eeqa}{\end{eqnarray}}

\begin{document} 
\begin{titlepage} 
\rightline{\vbox{\halign{&#\hfil\cr
&ANL-HEP-PR-97-09\cr 
&\today\cr}}}
\vspace{0.25in} 

\begin{center} 
 
{\large \bf ELECTROWEAK-SCALE EXCESS CROSS-SECTIONS  
IN THE SEXTET QUARK ``STANDARD MODEL''}

\medskip

Alan R. White\footnote{Work supported by the U.S. Department of 
Energy, Division of High Energy Physics, Contract\newline W-31-109-ENG-38}
\\ 
\smallskip
High Energy Physics Division, Argonne National Laboratory, Argonne, IL 
60439.\\ 

\end{center} 

\vspace{1in}
\begin{abstract} 

When the Higgs sector of the standard model is replaced by a flavor doublet
of color sextet quarks, dynamical electroweak symmetry breaking occurs, with
the electroweak scale identified as the sextet chiral scale. Above the
electroweak scale QCD is almost scale-invariant, leading to the high-momentum
enhancement of both the sextet chiral condensate and instanton interactions. 
The instanton interactions combine with the sextet condensate to produce
electroweak scale chirality violating processes in both the triplet and
sextet sectors. If the enhancement is sufficient, an electroweak scale top 
mass can be produced and the $\eta_6$ axion mass can be raised, potentially
providing a consistent description of both Strong CP conservation and
electroweak-scale CP violation. The instanton interactions 
also produce electroweak scale chirality violating scattering
processes for light quarks and additionally modify QCD evolution at the 
electroweak scale. Such effects could explain the excesses in the DIS
cross-section at HERA and jet cross-sections at the Tevatron.

\end{abstract} 

\end{titlepage}

%\vspace{.2in}

\mainhead{1. INTRODUCTION}

Despite the euphoria over the possible observation of a leptoquark, the 
possible discovery of supersymmetry etc., it is possible that 
what has been seen\cite{HE} at large $x$ and $Q^2$ at HERA is simply an
excess in the cross-section compared to the standard model prediction. Since 
the scales involved are the same, it
is also quite likely that this is essentially the same phenomenon as the large 
$E_T$ excess in the jet cross-section observed\cite{CDF} by CDF at the 
Tevatron. If this is the case, these phenomena could be
a crucial pointer towards the
improvement of the standard model. It is the current wisdom that the
unresolved problems of the model require a significant enlargement
of the $SU(3)\otimes SU(2) \otimes U(1)$ gauge theory. Supersymmetry is
overwhelmingly advocated as the main ingredient and many theorists (and
experimenters) are persuaded that viable non-supersymmetric alternatives for
the Higgs' sector are essentially non-existent. In this paper we will
summarize the virtues of replacing the Higgs sector of the standard model by
a flavor doublet of color sextet quarks. This leaves the gauge group
structure intact\footnote{This is not strictly true in that fermion masses
are assumed to arise from non-renormalizable four-fermion couplings which
ultimately should result from a larger theory.} and does not involve
supersymmetry. Indeed the standard model is modified in a very minimal
manner. Nevertheless, there are major consequences, including the potential
resolution (in an aesthetic manner) of several of the ``unresolved
problems'' of the standard model, particularly CP violation. 

The deepest property of the sextet quark model(SQM) is that electroweak symmetry
breaking is a ``strong interaction effect'' with the QCD sextet chiral scale 
determining the ``electroweak scale'', i.e. the mass of the $W^{\pm}$ and 
$Z^0$. As a result both the electroweak and strong interactions are modified
in a correlated way above the electroweak scale. It seems possible that the
linkage of the electroweak scale to the strong interaction is the essential
feature of the short-distance excess cross-sections seen at Fermilab and at
DESY. Within the SQM, the longitudinal components of the $Z^0$ and 
$W^{\pm}$ are ``sextet pions'' and so inherit a set of QCD
chiral interactions. Most relevant at low-enery are the couplings to the SQM 
axion, the $\eta_6$. A-priori, we expect the (non-chiral) sextet 
dynamical scale to be at least several hundred GeV and probably even higher.
General perturbative QCD interactions linking the triplet and
sextet sectors will not be significant until well 
above this scale. 

QCD instanton interactions also link the two 
sectors. It is well-known that instanton interactions play a fundamental 
role in the low-energy effective lagrangian of QCD because they are the
source of the light quark U(1) axial charge violation\cite{gth}. 
At the electroweak scale, non-perturbative instanton effects are very small in
conventional QCD. However, the presence of the sextet sector implies that at
the lowest sextet infra-red scale, i.e. the electroweak scale, instanton
axial charge violation (i.e. left-right transitions) in the light quark
triplet sector will mix with that in the sextet sector. Hence chirality
violation in the light quark sector may re-emerge. If it does the
amplitudes involved will grow in significance as the dynamics above the
electroweak scale is exposed. Such interactions are clear candidates for new
dynamics that appears to be emerging at the electroweak scale but which, in
many respects, is indistinguishable from lower-energy physics - as seems to
be the case for the cross-section excesses observed. 

Instanton interactions link infra-red and ultraviolet phenomena and so
do not scale in a straightforward manner. In the SQM, QCD becomes an almost 
scale-invariant theory above the sextet dynamical scale and high-momentum
effects are strongly enhanced. As a result interactions involving color
instantons at, or smaller than, the (inverse of) the 
electroweak scale are much stronger than in conventional QCD. These 
are the interactions that can produce observable axial charge violation 
effects at the ``infra-red'' electroweak scale. (We should emphasize that
while large chirality violating amplitudes are a measure of the instanton
effects, chirality conserving amplitudes are also enhanced by their
presence.) It is difficult to 
quantify this phenomenon but if it is as large as we assume then there are many 
interesting consequences. In particular, the electroweak scale of the top
quark mass is understood as a QCD instanton effect which is linked to a
number of other chirality violating effects at the same scale. Amongst such
effects is a significant increase in the mass of the $\eta_6$ axion,
possibly providing a novel resolution of the problems of CP violation.
According to a very crude estimate, if the top quark mass is due to
instanton effects as we describe then the $\eta_6$ also aquires a mass that
is at, or not far below, the electroweak scale. 

In the elementary one instanton approximation, which is all we discuss in
any detail in this paper, the quark mass spectrum below the electroweak scale
is essentially the inverse of the bare spectrum at the higher momentum scale
where the instanton interactions are strong. The anomalously high value of
the top quark mass is then a consequence of an anomalously low mass
at the higher scale. Correspondingly light quarks should aquire
momentum-dependent dynamical masses that increase strongly at
the electroweak scale. These masses are the simplest example of left-right
transition amplitudes and their existence is a general
ingredient in the occurrence of large instanton contributions to parton
model processes involving left-right quark transitions. Although they modify
the theory in only a minimal way, such transitions are absent, of course, in
conventional perturbative QCD. We expect similar effects in all electroweak
scale processes, but they should be most visible in the simple large $Q^2$
quark parton interaction seen at HERA. 

Although we expect them to be a small quantitative contribution,
we use the dynamical masses to obtain a qalitative 
idea of the general properties of excess cross-sections due to instanton 
effects. In particular, they naturally preserve the
angular distribution of the jet cross-section at the Tevatron, perhaps the
most surprizing feature of the excess! It is also clear that such effects
will be qualitatively similar to, and easily confused with, those obtained
by modifying parton distributions. 

It is many years since Marciano first proposed\cite{wm} ``higher-color'' as a 
special version of the technicolor idea for electroweak symmetry breaking. 
Not long after this\cite{arw} we were drawn to the proposal by the
realization that, if a flavor doublet of color sextet quarks played the role
in electroweak symmetry breaking suggested by Marciano, simultaneously the
Regge limit of QCD might be solved in a manner that we had come to believe
was essential\cite{arw1}. Since that time we have constantly advocated the
idea\cite{bww}. We have used it in attempts to explain both earlier
``new physics'' effects\cite{arw2}, which (apparently) have not withstood
the test of time, and high-energy cosmic ray physics\cite{arw3}. We have
also incorporated it in our general understanding\cite{arw1} of how Critical
Pomeron behavior\cite{cri} is obtained in QCD. 

We were persuaded of the virtues of the SQM fom the outset because of its
significance for the QCD Pomeron\footnote{In fact recent results from H1 on
DIS diffractive scaling violations\cite{H1} appear to support\cite{arw4} our
Super-Critical Pomeron solution of QCD that anticipates the appearance of
the Critical Pomeron at energies above the sextet scale.}  but we have only
gradually realized some of the additional theoretical and experimental 
implications. In particular we have slowly understood how 
the model is fundamentally different from technicolor theories. Or perhaps 
we should say that the SQM naturally contains all the features that 
are known to be necessary to avoid problems in the technicolor framework and 
it also has additional features. This 
is very important since it is well-known that any simple version of 
technicolor is ruled out by a number of experimental results. The
``technicolor'' gauge theory of the SQM is not simply a higher-scale analogue
of conventional QCD. Rather it is a very different version of QCD itself in
which $\alpha_s$ ``walks'', i.e. (almost) does not evolve, and the dynamics
is dominated by (almost) scale-invariant instanton interactions.

The importance of ``walking'' and the resultant ``high-momentum enhancement'' 
has been emphasized much in technicolor studies\cite{ta}. However, we believe 
it is the presence of the enhanced QCD instanton 
interactions coupling the triplet and sextet sectors which 
allows many of the problems of technicolor to be resolved by the SQM. 
There is no direct analogue for such interactions in technicolor theories.
As we have already noted, these same interactions may produce small, but
perhaps detectable, deviations from the standard model at current
accelerators. In the hope that the experimental evidence now appearing may
finally start to support this, and perhaps other, aspects of the SQM, 
we devote the first part of this paper to summarizing the model and some of 
the dynamical ingredients. We then discuss the instanton interactions, 
giving arguments for their enhancement and elaborating on the dynamical mass 
effects and other interactions. We also say what we can about the missing 
component of the SQM, the $\eta_6$, that may well be discovered at LEP or 
the Tevatron. The $\eta_6$ is the axion of the SQM but it also resembles the 
Higgs particle of the standard model. Instanton effects dominate it's
production and decay properties but if it is seen experimentally it could 
well be initially confused with the Higgs. Finally we outline how the
results from the Tevatron and HERA can be explained in terms of instanton
effects. Our explanations may be frustratingly qualitative and
oversimplified. Unfortunately we are unable to do much better. A partial
excuse is that instanton interactions are notoriously difficult to handle
quantitatively. (It is twenty years since 't Hooft first pointed out that
hadronic scale instantons solve the U(1) problem and provide an $\eta'$
mass, but even today this remains only a semi-quantitative
understanding\cite{ku}.) We finish with some additional ramifications of the
SQM. 

It is interesting to note that sextet quarks occur commonly in grand unified
and string models. They are sometimes referred to as quixes\cite{gg}. It is 
perhaps necessary to give arguments for why they should not occur, rather
than why they should! The most important point, however, is that within QCD
they are naturally associated with physics at the electroweak scale. We
should emphasize that at high enough energy the {\it ``new physics'' in the
SQM is not confined to small cross-section processes} which could affect
only rare events in very high-energy accelerators or very early universe
dynamics. It could be a major component of modern day cosmology and
astrophysics and should produce large cross-section physics, at the LHC for
example. On the theoretical side we might also point out that the infra-red
fixed point, (close to) zero $\beta$-function, version of QCD contained
in the massless SQM, which we argue is controlled by topological (instanton)
dynamics may eventually be amenable to solution via topological
methods\cite{ew}. It might, perhaps, be anticipated that such a theory is
essentially perturbative (and conformally symmetric) and so does not have a
confinement 
spectrum. In fact our Regge limit results\cite{arw4} imply that the
instanton interactions do produce confinement in the SQM version of QCD. We
also believe that this version of QCD may provide an attractive realization of
how, in the Regge limit, a conformally symmetric approximation - the BFKL
Pomeron\cite{lip} - underlies\cite{bpz} a second-order phase transition
theory - the Critical Pomeron\cite{cri}. 

\mainhead{2. ELECTROWEAK SYMMETRY BREAKING }

We add to the Standard Model (with no scalar Higgs sector), a massless 
flavor doublet $Q\equiv(U,D)$ of color sextet quarks with the {\em usual
quark quantum numbers}, except that the role of quarks and antiquarks is
interchanged. For the $SU(2)\otimes U(1)$ anomaly to be cancelled there must
also be other fermions with electroweak quantum numbers added to the
theory\cite{wm,bww}, but we will not consider the possibilities here. The
description of electroweak symmetry breaking parallels the conventional
technicolor treatment. 

We consider first the QCD interaction of the massless sextet quark sector.
There is a $U(2)\otimes U(2)$ chiral flavor symmetry. We anticipate that, as 
we discuss further below, a condensate develops\footnote{For color sextets,
two flavors are sufficient\cite{sm} for the 't Hooft anomaly condition to
require that chiral symmetry breaking accompanies confinement.} which breaks
the axial symmetries spontaneously and produces four massless pseudoscalar
mesons (Goldstone bosons), which we denote as $\pi^+_6,\;\pi^-_6,\;\pi^0_6$
and $\eta_6$, in analogy with the usual notation for mesons composed of $u$
and $d$ color triplet quarks. 
As long as all quarks are massless, QCD is necessarily $CP$ conserving in
both the sextet and triplet quark sectors. Therefore, in the massless 
theory we can, in analogy with the familiar treatment of flavor isospin
in the triplet quark sector, define sextet quark vector and axial-vector
currents $V^{\tau}_{\mu}$ and $A^{\tau}_{\mu}$ 
$$
V^{\tau}_{\mu}~=~\bar{Q}\gamma_{\mu}\tau Q~, ~~~~~
A^{\tau}_{\mu}~=~\bar{Q}\gamma_5 \gamma_{\mu}\tau Q~, 
\auto\label{cur}
$$
which are ``isotriplets'' under
the unbroken $SU(2)$ vector flavor symmetry together with singlet currents
$v_{\mu}$ and $a_{\mu}$ 
$$
v_{\mu}~=~\bar{Q}\gamma_{\mu} Q~, ~~~~~
a_{\mu}~=~\bar{Q}\gamma_5 \gamma_{\mu} Q~. 
\auto\label{cur1}
$$

The pseudoscalar mesons couple ``longitudinally''
to the axial currents, i.e. 
$$
<0|A^\tau_{\mu}|\pi^{\tau}_6(q)>~
\centerunder{$\sim$}{\raisebox{-4mm}{$q_{\mu} \to 0$}}~~ F_{\pi_6}q_{\mu}
\auto\label{lon}
$$
$$
<0|a_{\mu}|\eta_6(q)>~
\centerunder{$\sim$}{\raisebox{-4mm}{$q_{\mu} \to 0$}} ~~F_{\eta_6}q_{\mu}
\auto\label{lon1}
$$
Note that, for reasons that will become apparent later, we distinguish 
$F_{\eta_6}$ from $F_{\pi_6}$. If we define right-handed and left-handed
currents 
$$ 
2R^{\tau}~=~A^{\tau}+ V^{\tau},~~~2 L^{\tau}~=~   A^{\tau} - V^{\tau},~~~ 
2r~=~a +v  ,~~~2 l~=~a - v,~~
\auto\label{cur2}
$$
then since the vector currents remain conserved, we also have  
$$
<0|R^\tau_{\mu}|\pi^{\tau}_6(q)>~~\sim~~ <0|L^\tau_{\mu}|\pi^{\tau}_6(q)>~~
\centerunder{$\sim$}{\raisebox{-4mm}{$q_{\mu} \to 0$}}~~ {1 \over 2}~
F_{\pi_6}q_{\mu}
\auto\label{lon2}
$$
and
$$
<0|r_{\mu}|\eta_6(q)>~~~\sim~~ <0|l_{\mu}|\eta_6(q)>~~
\centerunder{$\sim$}{\raisebox{-4mm}{$q_{\mu} \to 0$}} ~~{1 \over 2}
~F_{\eta_6}q_{\mu}
\auto\label{lon3}
$$

We consider next the coupling of the electroweak gauge fields to the sextet
quark sector. The massless $SU(2)$ gauge fields $W^{\tau}_{\mu}$ couple to the
isotriplet sextet quark currents in the standard manner, that is 
$$
{\cal L}_I~=~gW^{\tau\mu}L^{\tau}_{\mu} ~=~
gW^{\tau\mu}\Bigl(A^{\tau}_{\mu} - V^{\tau}_{\mu} \Bigr)
\auto\label{cou}
$$
The U(1) hypercharge field $Y_{\mu}$ couples via
$$
{\cal L}'_I~=~ g'Y^{\mu}(R^Y_{\mu} - L^Y_{\mu})
\auto\label{cou1}
$$
where 
$$
R^Y~=~ r + R^0,~~~~~~L^Y~=~ l 
\auto\label{cou2}
$$

There are two well-known and very important features of the above hypercharge
couplings. The first is that 
$$
J_{em}~=~R^Y - L^Y - L^0
\auto\label{cou3}
$$
is a conserved vector current which does not couple to any of the Goldstone 
bosons. It is, of course, the electromagnetic current. Secondly, the current 
that couples to the hypercharge field, i.e.
$$
R^Y - L^Y~,
\auto\label{cou4}
$$
contains only the $\pi^0_6$ and not the $\eta_6$. Note, however, that
the right and left-handed hypercharge currents both couple to the $\eta_6$, 
i.e. 
$$
<0|R^Y_{\mu}|\eta_6(q)>~~\sim~~ <0|L^Y_{\mu}|\eta_6(q)>~~
\centerunder{$\sim$}{\raisebox{-4mm}{$q_{\mu} \to 0$}}~~ {1 \over 2}~
 F_{\pi_6}q_{\mu}~.
\auto\label{lon4}
$$
Combining (\ref{lon}) with (\ref{cou}), it follows that the
$\pi^+_6,\;\pi^-_6$ and $\pi^0_6$ are ``eaten'' by the usual combinations of 
the $SU(2)$ and U(1) gauge bosons and they respectively become the third
components of the $W^+,\;W^-$ and $Z^0$. 

It is straightforward to construct\cite{bww} a
gauge-invariant\footnote{This lagrangian will be equivalent to the unitary 
gauge formulation of the standard model.} $SU(2)\otimes U(1)$ gauged
chiral effective lagrangian for the sextet pion sector of the form 
$$
{\cal L}_{chiral}~=  ~{1 \over 4}~ F_{\pi_6}^2Tr({\cal D}_{\mu}U
{\cal D}_{\mu}U^{\dagger})
\auto\label{eff}
$$
where $U~=~exp({i~ \tau \cdot  \pi_6 /F_{\pi_6}})$ is the chiral field.
After triplet quark and lepton couplings are added, the 
chiral condensate produces the usual tree-level standard model
lagrangian for the electroweak sector. This lagrangian will correctly
describe $W$ and $Z$ interactions in the ``infra-red'' region, i.e. momenta
of order $m_Z$ (or $m_W$) with the $Z^0$ (or $W^{\pm}$) close to mass-shell.

We conclude that QCD chiral symmetry breaking generates
masses for the $W^+,\;W^-$ and $Z^0$ with $M_W\sim g\;F_{\pi_6}$. 
$F_{\pi_6}$ is {\em a QCD scale} defined by (\ref{lon}). It is reasonable to
anticipate that the relative scales of triplet and sextet chiral symmetry
breaking differ only because of the different Casimirs. In this case 
we can expect the ``Casimir Scaling'' rule\cite{wm,bww} to be approximately
valid, i.e.
$$
C(6)\alpha_s(F^2_{\pi_6})~\sim~C(3)\alpha_s(F^2_{\pi})
\auto\label{cas}
$$
where 
$$
C(6)/C(3) ~=~5/2~.
\auto\label{cas1}
$$ 
$C(6)$ and $C(3)$ are sextet and triplet casimirs respectively.
Given the standard low-energy evolution of $\alpha_s~$,
(\ref{cas}) is clearly consistent with $F_{\pi_6}\sim 250$ GeV! 
Also, since we are completely restricted to a flavor doublet if we insist on 
asymptotic freedom for QCD, the form of the
symmetry-breaking is automatically equivalent to that of an $SU(2)$ Higgs
sector and so 
$$
\rho=~M^2_W/M^2_Zcos^2\theta_W~=~1
\auto\label{sym}
$$
as required by experiment. 

To write a full low-energy 
``standard model'' lagrangian we should 
integrate out the sextet sector of QCD completely, but keep 
the full QCD lagrangian to describe the interaction of color triplet
quarks. At the level that all fermions are massless, the ``standard model'' 
electroweak remnants of the sextet sector are then 
the electroweak interaction (\ref{eff}), together with the ``axion 
coupling'' of the $\eta_6$ to QCD gauge fields that we discuss in the next
Section. (As we noted, the $\eta_6$ does not couple directly to the 
$SU(2)\otimes U(1)$ gauge fields.)  
There will also be QCD induced interactions amongst the $Z^0$, 
$W_{\pm}$, $\eta_6$ and the triplet quarks that will be described in 
Section 5. 

To produce fermion masses we will follow the standard technicolor path and
introduce gauge-invariant, but non-renormalizable, lepton/sextet and
triplet/sextet four-fermion couplings
(that should ultimately be traceable to a larger unifying gauge group).
When combined with the sextet quark condensate, such couplings provide 
lepton and triplet quark ``bare masses'' via 
\beq
\label{ff}
{c \over {\Lambda}^2}~ Q\bar{Q}f\bar{f}~~\to~~~{c~\VEV{Q\bar{Q}} \over 
{\Lambda}^2} f\bar{f} ~~\to~~ m_f f\bar{f}
\eeq 
where we expect $c$ to be a dimensionless coupling constant and $\Lambda$ to
be a scale (presumably the mass of a gauge boson in the unifying gauge
group). In our discussion $\Lambda$ will simply be the cut-off at which the 
physics we discuss is modified by new physics. It will be important 
that {\it the bare masses given by (\ref{ff}) are defined above the 
electroweak 
scale.} Because of the instanton interactions that we discuss in Section 
5 the mass spectrum below the electroweak scale will be radically different 
from the bare mass spectrum.

It is essential that the chiral symmetry involved in electroweak symmetry 
breaking is broken only dynamically, so that the Goldstone 
bosons produced have strictly zero mass before mixing with the electroweak
gauge fields. The presence of the four-fermion couplings requires that this 
be a combined triplet/sextet chiral symmetry, the left-handed part of which 
is the SU(2) gauge symmetry. Assuming that the sextet chiral scale is indeed 
much bigger than the triplet scale, as implied by (\ref{cas}) and as we 
discuss further in Section 4, the ``sextet pions'' will 
contain just a small component of the normal triplet pions. For simplicity we 
will continue to treat the sextet chiral symmetry as entirely responsible 
for electroweak symmetry breaking. Correspondingly, we will treat the triplet 
chiral symmetry as explicitly broken by the couplings (\ref{ff}) while the 
sextet symmetry is not. In this case the sextet quarks should
have only the dynamical mass discussed in the next Section, with the 
corresponding condensate producing ``bare'' triplet quark masses via 
(\ref{ff}).

In summary, the SQM contains a (very particular) version of technicolor
symmetry breaking which fits experiment and has the attractive property that
{\it the electroweak scale is naturally explained as a second $QCD$ scale.} 

\mainhead{3. THE $\eta_6$ AND CP VIOLATION}

Before discussing QCD sextet quark dynamics we first describe the role of 
the $\eta_6$ in the SQM. Since it is not involved in generating masses for
the electroweak gauge bosons, the $\eta_6$ remains as a Goldstone boson
associated with the sextet $U(1)$ axial chiral symmetry and is massless
before triplet quark masses are added to the theory. It couples to the $QCD$
color anomaly via the sextet-quark triangle anomaly and gives a low-energy 
effective lagrangian of the form 
\beq
\label{laga}
{\cal L}~=~{\cal L}_{QCD}+\tilde{\theta}\frac{g^2}{32\pi^2}
F\tilde{F}+
\frac{\eta_6}{F_{\eta_6}}\frac{5 g^2}{64 \pi^2}
F\tilde{F}+\cdots
\eeq
where ${\cal L}_{QCD}$ is the usual QCD lagrangian for
the gauge and triplet quark sectors and, in a conventional notation,
$\tilde{\theta}=\theta +\arg\det m_3$, where $m_3$ is the triplet quark mass
matrix. 

The $\eta_6$  is {\it an axion} very like that originally envisaged\cite{ww}
as producing  Strong $CP$ Conservation via the Peccei-Quinn
mechanism\cite{pq}. Within the lagrangian (\ref{laga}) an appropriate shift 
in the $\eta_6$ field absorbs the 
$CP$-violating $\tilde{\theta}$ term and a sufficient condition for 
$CP$-conservation is that a minimum of the axion potential occurs at
$\hat{\theta}=0$ (where $\hat{\theta}=\theta + \arg \det m_3+ 5\VEV{\eta_6}/
2 F_{\eta_6}$). The axion mass is generated by the curvature of the potential
at the minimum. If all of the relevant dynamics involves only the
normal $QCD$ scale $\Lambda_{QCD}$, and triplet quark masses, this mass is
inevitably of O($\Lambda^2_{QCD}/ F_{\eta_6}$) and hence very
small\cite{pec}. 

In conventional QCD, instanton interactions give only a very small indirect
contribution to the 
axion mass. The major contribution comes from the direct effects of quark
masses. As we discuss further in Section 5, and as we emphasized in the
Introduction, within the SQM instanton interactions are strongly enhanced 
above the electroweak scale. At this scale, the only contribution of the 
triplet quarks is via their bare masses - which badly break the U(1) 
symmetry associated with the $\eta_6$ axion. As a result, the instanton
dynamics relevant for the $\eta_6$ can be regarded as entirely contained 
within the sextet sector, but analagous to that of the 
triplet sector at the usual triplet scale. Since the $\eta_6$ plays the 
analagous role to that of the $\eta'$, it is clear that there is an additional 
large contribution to the $\eta_6$ mass. Once again the mechanism involved
is close to that originally envisaged\cite{hol} within a more general
technicolor theory but operates with particular strong effect in the SQM.
Sextet instanton interactions generate a large number of $\eta_6$ vertices
which, when instanton and anti-instanton interactions are added, all contain
a factor (or factors) of $\cos [\tilde{\theta}+ \VEV{\eta_6}]$. Therefore,
provided there are no additional sextet sector interactions, 
the axion potential generated naturally retains the $CP$-conserving
minimum at $~\tilde{\theta}+\VEV{\eta_6} =0$ while simultaneously producing
a large contribution to the $\eta_6$ mass. That the instanton $\eta_6$
mass can be large is another important ``special'' aspect of the SQM in
comparison with general technicolor models. Again it is well-known that a
Peccei-Quinn axion with the standard mass has been ruled out by
experiment\cite{pec}. 

From our current perspective, the overall magnitude of the triplet/sextet
four-fermion couplings (\ref{ff}), or equivalently the overall magnitude of 
the fermion bare mass spectrum, can be regarded as a 
parameter which can be smoothly raised from zero to its physical value.
During this variation the $\eta_6$ mass will also go from zero to its physical
value. When the $\eta_6$ mass is at zero all of the additional
sextet sector physics remains above the sextet chiral scale. Therefore
the triplet sector axion status of the $\eta_6$ will be clear initially and 
will be preserved during the variation. Consequently the theory will stay at 
the CP conserving minimum. Strong CP will be conserved by the triplet quark
sector even if the $\eta_6$ mass is raised to the electroweak scale, as we
will argue in Section 5. We will also discuss possible decay modes 
of the the $\eta_6$ in Section 5. 

Focussing now on the $CP$ properties of the full sextet sector above the
electroweak scale, we note that 
the Peccei-Quinn argument is inapplicable since we can not write a
lagrangian of the form (\ref{laga}). That is we can not write a lagrangian 
involving both the $\eta_6$ and the gluon field to describe general sextet 
quark 
interactions. If the gluon field is to be present, then we must use the full
$QCD$ lagrangian, written in terms of elementary fields, for the combined
triplet and sextet sectors. This clearly has no axion. {\it Because there is
no axion}, QCD interactions above the electroweak scale will naturally be
{\it Strong $CP$-violating}. In other words there can be a non-zero  
``$\theta$-parameter'' in the full sextet scale QCD lagrangian. Also, the 
mixing of the chiral symmetries discussed in the last Section implies that
the familiar triplet quark mesons (i.e. the pions and kaons) will contain a 
small admixture of sextet quark states which could very well provide their
$CP$ violating interactions. Therefore {\it electroweak scale $CP$-violation
may actually be ``Strong $CP$-violation'' within the SQM.} 

Replacing the Higgs sector with a sextet quark sector removes one of the
most unsatisfactory features of the standard model. The low energy theory
has $CP$-violation which is not wanted and is removed in an ugly manner. At
the electroweak scale $CP$-violation is required and has to be added back in
to the model in an equally ugly manner. Both low-energy conservation and 
high-energy violation of $CP$ are natural consequences of the SQM.

\mainhead{4. THE $\beta$-FUNCTION AND WALKING COLOR}

We consider now ``perturbative'' chiral dynamics within the QCD sector of
the SQM. If we write the QCD $\beta$-function in the form 
\beq
\label{beta}
\beta(\alpha_s) = - \beta_0 \alpha_s^2(q)/2\pi~ -~ 
\beta_1 \alpha_s^3(q)/8\pi^2 + ~~...
\eeq
then for six color triplet flavors the normal two-loop calculation gives 
\beq
\label{six}
\beta_0 = 11 - 2n_f/3~ = 7,~~~\beta_1 = 102 - 38n_f/3~= 26 ~.
\eeq
The corresponding $\beta(\alpha_s)$ is shown in Fig.~4.1(a).
When the two sextet flavors are included we obtain\cite{tar}
\beq
\label{sex1}
\beta_0 = 7 - 4T(R)n^6_f/3~ = 7 - 4(\frac{5}{2})2/3~ = 1/3,
\eeq
and
\beq
\label{sex2}
\beta_1 = ~26 - 20T(R)n^6_f - 4C_2(R)T(R)n^6_f~ =~26 - 100 -66\frac{2}{3} 
~=~-140\frac{2}{3}   
\eeq
where $T(R) = C(R)/C(3) = 5/2$ and $C_2(R) = 10/3$ for sextet quarks.
The corresponding SQM $\beta$-function is shown in Fig.~4.1(b). It 
is (just) asymptotically-free and also has an
infra-red fixed point at 
$$
\alpha_s~\approx ~ 1 /  34
\auto\label{as}
$$
(There is a sense in which this can be argued to be present to all 
orders\cite{bz}). In addition, 
between the ultra-violet and infra-red fixed points the $\beta$-function
remains very small ($ <~10^{-6}$). As a result the massless theory
evolves only very slowly and is 
almost scale-invariant. 

We would like to have at least a qualitative idea of how the massive SQM
behaves from immediately above the electroweak scale up to the asymptotic
energies where the massless $\beta$-function is relevant. We assume that
for this we can 
use an effective $\beta$ function which describes the evolution of some 
appropriately defined $\alpha_s$ and which interpolates between Fig.~4.1(a)
and Fig.~4.2(b). This is what we have shown in Fig.~4.1(c).

\begin{center}

\leavevmode
\epsfxsize=4.5in
\epsffile{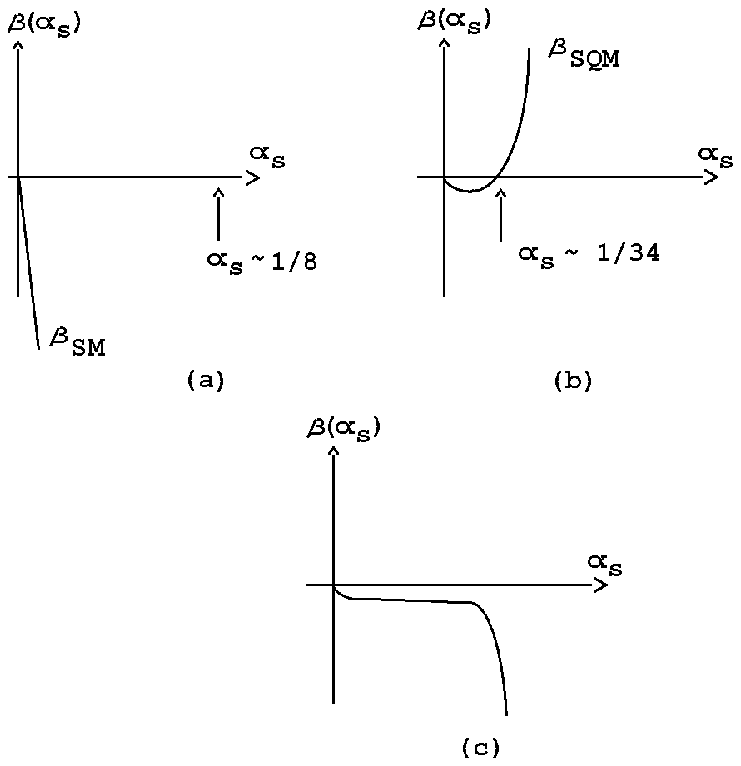}

Fig.~4.1 (a) The Standard Model QCD $\beta$-function (b) The QCD 
$\beta$-function 
for the Massless SQM (c) An Effective QCD $\beta$-function in the Massive SQM.
\end{center}
The extended 
shoulder reflects the presence of the infra-red fixed point in the massless
$\beta$-function. According to Fig.~4.1(c), $\alpha_s$ will dramatically 
stop it's conventional evolution at some scale and then will evolve 
extremely slowly from this point on. Of course, the precise definition of
$\alpha_s$ will also be important. It is possible that, as we suggest in 
Section 6, it can be defined via jet cross-sections in such a way that the 
evolution stops already at the electroweak scale where the cross-section 
excess begins to appear. 

Working with the $\beta$-function of Fig.~4.1(c), we have circumstances very
close to those originally envisaged for the ``walking technicolor'' idea,
which we adapt to ``walking color'', and the resulting ``high-momentum
enhancement''. The major ingredients\cite{app} 
are the existence of a small $\beta$-function (approximated by a constant)
and the assumption that the linearised Dyson-Schwinger ``gap
equation'' for the quark dynamical mass $\Sigma(p)$,
gives a semi-quantitive description of the dynamics of short-distance chiral
symmetry breaking. Because of the additional role played by instantons we 
will apply the ideas involved in a slightly different manner. However, we 
first give a very brief and elementary summary of the ideas, as they are 
usually applied. 

As we noted in the last Section, only the triplet quarks will have a bare 
mass. Both the triplet and the sextet quarks will have a dynamical 
mass. The dynamical fermion mass $\Sigma(p)$ is defined by 
writing the full inverse fermion propagator in the form
$$
S^{-1}(p)~=~\st{p}(1+A(p)) - \Sigma (p)
\auto\label{dym}
$$
$A(p)$ plays a relatively insignificant role, it simply contributes to
wave function renormalization at large momentum. $\Sigma(p)$ plays
an important role and will figure prominantly in our discussion. 
The Dyson-Schwinger equation it satisfies can be
truncated to give the (zero bare mass) linearized approximation 
$$
\Sigma (p)~=~{3 C_2(R) \over 4\pi} \int_{|k| < \Lambda}~d^4k~ 
{\alpha_s((k-p)^2) 
\over k^2(k-p)^2}~ \Sigma(k)
\auto\label{dse}
$$
which is illustrated in Fig.~4.2. (The truncation introduces a 
gauge-dependence for $\Sigma(p)$. However, it can then be argued\cite{um}
that the dynamical features described below are gauge-invariant.)
We use the notation, in Fig.~4.2 and in later figures, that a line 
containing a $\hat{\Sigma}$ insertion represents $ \Sigma(p^2)/p^2$.
Lines without insertion, or lines attached to $\Sigma$ 
rather than $\hat{\Sigma}$, will represent the usual massless propagator.
\begin{center}

\leavevmode
\epsfxsize=3in
\epsffile{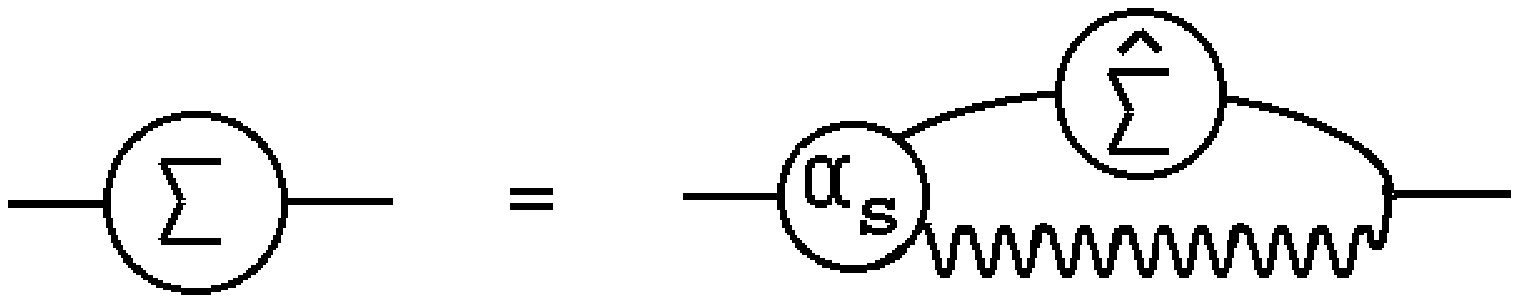}

Fig.~4.2 The Linearized Dyson-Schwinger Equation

\end{center}
For fixed $\alpha_s$ and finite $\Lambda$, this equation has a 
solution corresponding to spontaneous chiral symmetry breaking when 
$\alpha_s ~\centerunder{\raisebox{1mm}{$\scriptscriptstyle >$}}
{$\scriptscriptstyle \sim$}~ 
\alpha_c$, where $\alpha_c$ is determined by an equation of the form 
\beq
\label{chir}
C_2(R)~ \alpha_c = \mbox{constant}
\eeq
Since $C_2(R)$ is very similar in magnitude to $C(R)$, this is qualitatively
consistent with the Casimir scaling rule (\ref{cas}).

The solution of the gap equation for $\alpha_s \sim \alpha_c$ has the
form\cite{app} 
\beq
\label{sig}
 \Sigma (p) \sim ~\mu^2~(p)^{-1}
\eeq
where $\mu$ is the 
scale at which this behaviour sets in. When $\alpha_s$ is allowed
to ``walk'' according to a very small $\beta$-function (of the form of
Fig.~4.1(c)) the behavior (\ref{sig}) 
also appears and persists up to large momentum, before a faster falling 
asymptotic solution takes over. $\mu$ could be used to define 
the sextet quark constituent quark mass. 
It will presumably be 
not too far above the electroweak scale, although it will depend, of course,
on exactly how $\alpha_s$ is defined and the $\beta$-function of Fig.~4.1(c)
introduced. 

When the behavior (\ref{sig}) is inserted into the perturbative
formula\cite{app} for the high-momentum component of the sextet condensate
$\VEV{Q\bar{Q}}$, as illustrated in Fig.~4.3(a), 
we obtain a contribution 
\beq
\label{cond}
\VEV{Q\bar{Q}} \sim \int {d^4p \over p^2} \Sigma (p) 
\sim \int dp ~ p~ \Sigma(p) ~\sim \mu^2\Lambda
\eeq
where $\Lambda$ is the upper cut-off on the integral. 
In contrast, the 
corresponding perturbative formula\cite{app,pag} for the chiral constant
$F_{\pi_6}$ (which determines the $W$ and $Z$ masses as above) has a
high-momentum component of the form illustrated in Fig.~4.3(b), i.e.
\beq
\label{ps}
F^2_{\pi_6} \sim \int dp~p^{-1}~\Sigma^2(p)~~+~~...
\eeq
which is not enhanced by the behavior (\ref{sig}). 
\begin{center}

\leavevmode
\epsfxsize=5in
\epsffile{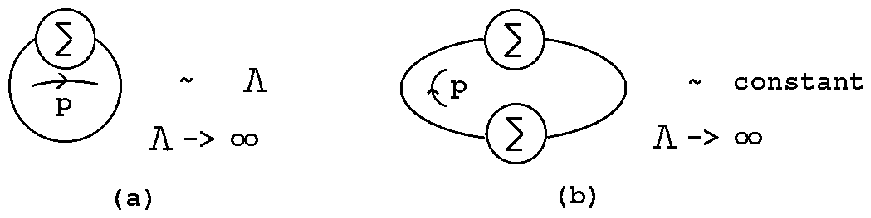}

Fig.~4.3 Integrals Giving (a) $\VEV{Q\bar{Q}}$ and (b) $F_{\pi_6}$ 

\end{center}
Therefore we can view $F_{\pi_6}$ as essentially produced directly by chiral
(i.e. electroweak scale) sextet interactions. We can anticipate a similar 
(order of magnitude) contribution to the condensate from the ``infra-red''
region, with the enhancement being a high-momentum effect.

So far we have identified $\Lambda$ as the scale at which further new 
(unification) physics appears. As we have already noted, this physics is 
required to produce the triplet/sextet four-fermion couplings generating 
triplet quark masses and so $\Lambda$ can also be identified with the 
inverse scale for the (bare) fermion couplings. From (\ref{ff}) the masses
have the order of magnitude \beq 
\label{mas}
m_q \sim ~\VEV{Q\bar{Q}}/{\Lambda}^2 ~\sim ~\mu^2 / \Lambda
\eeq
The scale difference between the condensate and the chiral decay constant 
produced by (\ref{cond}) and (\ref{ps}) is argued to be essential in 
technicolor theories to obtain reasonable values for
quark masses without inducing large flavor-changing neutral
currents\cite{app}. A-priori the two scales are identified and the problem 
with flavor-changing processes then arises. In our case the instanton 
interactions make the low energy evolution of the mass spectrum more 
complicated but the separation of scales is still important. 

The above analysis is surely oversimplified but it does illustrate the general 
idea. That $\alpha_s$ starts to ``walk'' in
the SQM immediately above the sextet chiral scale, naturally leads to a
large momentum range over which the ``perturbative'' interactions are almost 
scale invariant. This provides a high-momentum dominated chiral dynamics in
which loop integrals of the form Fig.~4.3(a) are enhanced while those of
the form Fig.~4.3(b) are not. 

\mainhead{5. ENHANCED INSTANTON INTERACTIONS}

In this Section we include the chiral dynamics exemplified by Fig.~4.3
in a discussion of QCD instanton interactions involving sextet and triplet
quarks. The two sectors are, of course, simply linked by gluon exchange. 
However, as we discussed in the Introduction, the
instanton interactions, provide a link that may appear already at the 
lowest ``infra-red'' scale of the theory rather than the higher ``dynamical''
sextet scale. 

The presence of an infra-red fixed-point in the massless QCD
$\beta$-function of Fig.~4.1(b) is very closely related to the absence of
infra-red renormalons in the Borel plane\cite{arw1,cjm}. The two properties
imply that, within the massless theory, the perturbation expansion is more
convergent and also\cite{gp} {\it  instanton interactions have no infra-red
scale divergences}. As a result it is not necessary to remove instanton
infra-red divergences with additional, ambiguous, non-perturbative gluon
condensates\cite{fd}. {\it Instanton interactions provide all the
non-perturbative physics of massless SQM QCD}. The consequence for the
physical massive SQM is that QCD instanton 
interactions should be well-defined down to the infra-red electroweak scale and
provide all of the non-perturbative physics down to this scale. Combining
this infra-red finiteness with the extremely slow evolution of $\alpha_s$,
color instanton interactions persist over a very wide scale range.
For almost all of our discussion we will consider these interactions in the
original semi-classical one-instanton approximation of 't Hooft\cite{gth} in
which the fermion zero modes give rise to a point-vertex. This vertex simply
represents the axial charge violation at zero momentum 
and does not give any accompanying momentum dependence. 

The large Casimir of sextet quarks leads to a surprizingly high-order one 
instanton interaction. In the massless theory, the singlet current 
$$ 
\eqalign{J^0_{\mu}~=~a^6_{\mu}-5a^3_{\mu}}
\auto\label{cons}
$$
is conserved in the presence of instantons (6 and 3 now denote
sextet and triplet currents respectively). 
The one instanton zero modes produce an interaction which, as illustrated in
Fig.~5.1, corresponds to the vacuum production of a left-handed(say) pair of 
each quark flavor, subject to the conservation law given by (\ref{cons}).
\begin{center}

\leavevmode
\epsfxsize=2in
\epsffile{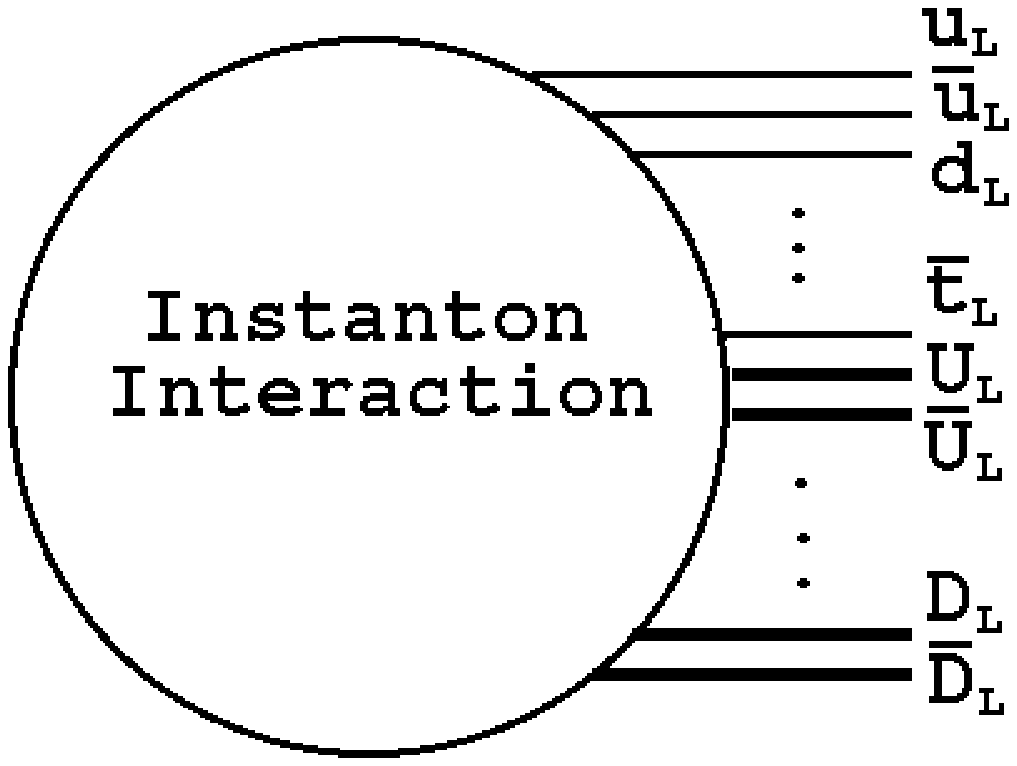}

Fig.~5.1 The One Instanton Interaction

\end{center}
This vertex involves one quark/antiquark pair of each triplet
flavor and five pairs of each sextet flavor i.e. a thirty-two, 
left-handed, fermion interaction of the form 
\beq
\label{inst}
\bar{u}_Ru_L\bar{d}_Rd_L\bar{s}_Rs_L\bar{c}_Rc_L\bar{b}_Rb_L\bar{t}_Rt_L
(\bar{U}_RU_L)^5(\bar{D}_RD_L)^5
\eeq
Of course, in addition to Fig.~5.1, this vertex also gives all the
interactions in which any number of the left-handed fermions are crossed
over to become incoming right-handed fermions. An anti-instanton gives the
corresponding right-handed interaction. The presence of such bizarre
interactions probably adds to the necessity for fermion condensates to
develop in order to give a sensible massless theory containing particles.
Conversely it is, in part, because so many pairs of sextet quarks are involved
that the interaction has a significant order of magnitude. 

Given the point instanton interaction (\ref{inst}) as an ``infra-red'', i.e.
electroweak scale, interaction we can close-up the sextet lines pairwise
with the sextet condensate, as illustrated in Fig.~5.2.
Equivalently we can close up lines with $\Sigma(p)$, then with the instanton 
interaction treated as a point vertex, each sextet loop
integral (over loop momenta $p$ as illustrated) gives directly the
condensate integral (\ref{cond}) of Fig.~4.3(a).
The result is, as illustrated, the usual QCD triplet interaction for
triplet quarks. 
$$
I ~\equiv~ c_I~
\bar{u}_Ru_L\bar{d}_Rd_L\bar{s}_Rs_L\bar{c}_Rc_L\bar{b}_Rb_L\bar{t}_Rt_L
\auto\label{trI}
$$
Although the treatment of the instanton as a point interaction over the 
whole momentum range is surely too simple, because of the approximate scale 
invariance of the theory we can anticipate that any momentum dependence is 
very gradual. Therefore we expect that the convolution integrals with 
$\Sigma(p)$ illustrated in Fig.~5.2, over an extended ``high-momentum'' 
region, significantly enhance the basic interaction. 
\begin{center}

\leavevmode
\epsfxsize=4.5in
\epsffile{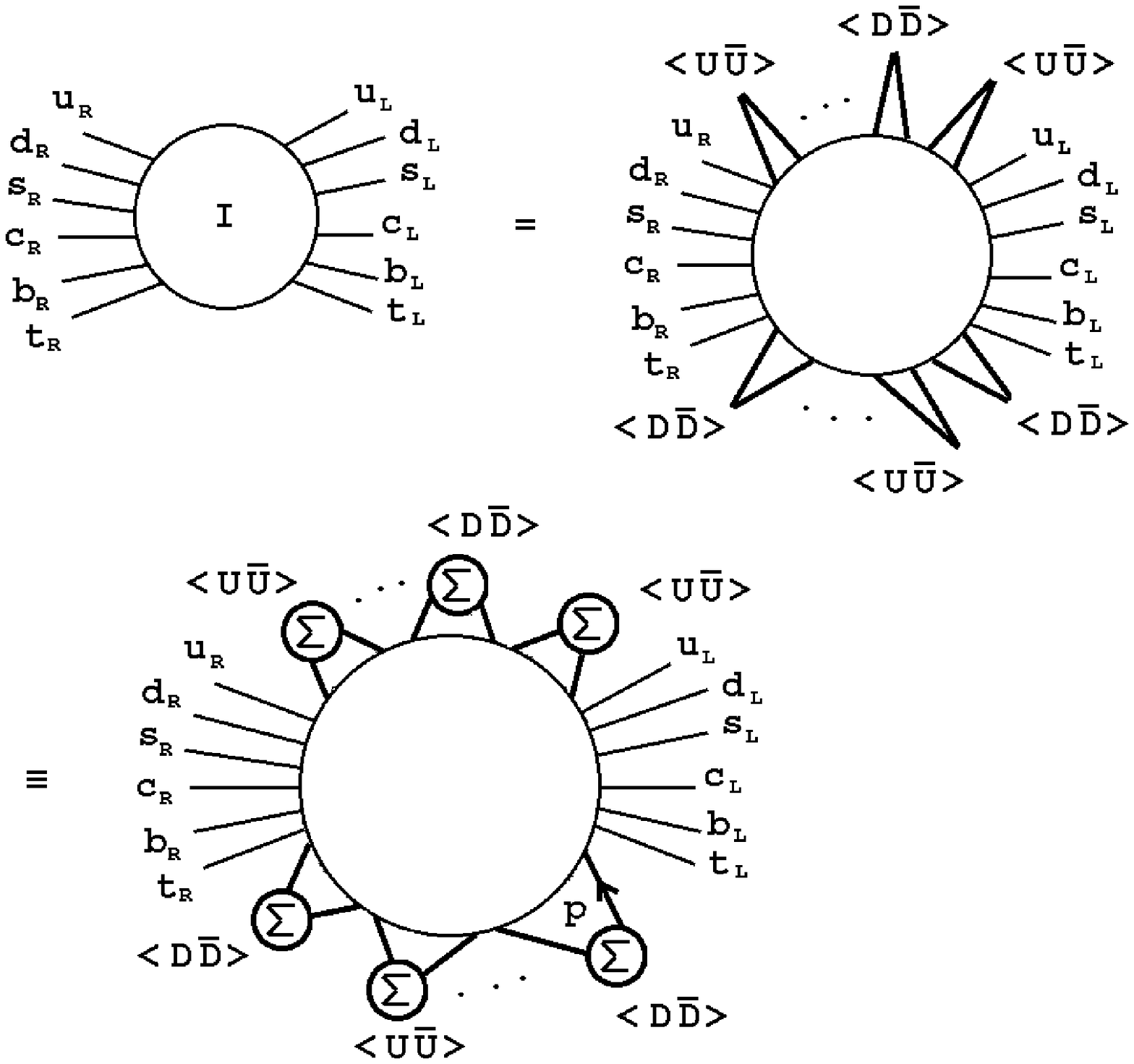}

Fig.~5.2 Closing Sextet Quark Lines in the Instanton Interaction.

\end{center}

Another major factor in the overall magnitude of $c_I$ 
is the usual exponential factor 
\beq
\label{sup}
c_I~\sim ~
[ \alpha_s ]^{-6}~
\exp{[-2\pi /\alpha_s]}
\eeq
In conventional QCD this factor not only suppresses the integrand
but, because of the running of $\alpha_s$, also cuts off the integration 
range. This leads to the conclusion that color instanton interactions are
very weak at the electroweak scale. If we take $\alpha_s ~\sim 1/8$ then 
$$
(\alpha_s)^{-6}~exp[ -2\pi /\alpha_s] ~~\sim ~~10^5~ exp[ - 50] ~~ \sim 
10^{-17}
\auto\label{efs}
$$
This is a big suppression factor, even for a twelve fermion interaction.
However, in our case we have a thirty-two fermion interaction and if we
consider this as resulting from an effective three-point coupling
$g_I$, raised to the thirtieth power, we obtain 
$$
g_I~ \sim~ \biggl( 10^{-17} \biggr)^{1/30} 
~~\sim ~~ (10)^{-1/2}~~\sim ~~1/3
\auto\label{bgn}
$$
which is not a small ``coupling constant''. If a large number of loop 
integrals are performed, each of which is ``high-momentum 
enhanced'', to obtain lower-order interactions, 
as we are effectively doing, it is clear that we need not obtain a small 
result. That we are actually dealing with a high-order multiple
interaction is also reflected in the enormous magnitude of the 
numerical factor that multiplies (\ref{sup}). A rough estimate of this 
factor is \cite{gth}
$$
(5!)^2 2^{58} {\pi}^{38} \sim 10^{40}~.
\auto\label{est}
$$
At first sight this immediately justifies our claim that there is a large 
effect. Unfortunately the factor (\ref{est}) is highly regularization 
procedure dependent\cite{gth}. Although this should be less significant,
given the infra-red finiteness discussed above. Of course, a major part of
this factor simply compensates for the factors of $(2\pi)^{-4}$ that accompany
the loop integrals involved in Fig.~5.2. In fact we believe the estimate
(\ref{bgn}), combined with the argument that both the instanton interaction and
perturbative self-interactions are high momentum enhanced, is sufficient to
ensure that large instanton effects will be obtained compared to 
conventional QCD. 

Actually the situation may be similar to the discussion of instanton physics 
at the triplet chiral scale. A qualitative idea of the physics involved can
be obtained from a discussion of the one instanton interaction. However, 
in part because the order of magnitude of this vertex can be made to vary
by changing the regularization procedure used, a more complete
treatment including, at least, multi-instanton effects is necessary to
obtain physically meaningful orders of magnitude. The ``instanton density''
then becomes a relevant parameter, in addition to the instanton size. This
is a problem which has not been unambiguously solved in conventional triplet
quark QCD. It may even be that, in analogy with discussions of chiral
dynamics at the triplet scale, collective instanton effects have to be
considered\cite{dp}. On this basis we shall try to arrive 
at a consistent picture for the relative size of the various interactions
and leave the overall magnitude to be determined phenomenologically. 
We will assume that we can work with one scale $\Lambda_I$ for the interactions 
which is essentially determined by the maximum size of contributing 
instantons. We assume that 
$$
 \mu ~~\centerunder{$<$}{\raisebox{-2mm}{$\sim$}}~~\Lambda_I  ~~< 
~~\Lambda~.
\auto\label{sca1}
$$

Consider now some of the processes to which the triplet interaction I 
contributes. If we close up all but one of the triplet lines with a bare 
mass we obtain a contribution to the dynamical quark mass matrix 
$\Sigma_3$. This is illustrated in Fig.~5.3 by the diagonal contribution to the 
top quark dynamical mass. 
In this case, again ignoring any momentum dependence of the instanton 
interaction, each loop integration has a ``large momentum'' contribution
(large here meaning $p \centerunder{$>$}{\raisebox{-2mm}{$\sim$}}
\sim \Lambda_I$) of the form
$$
m^0_f ~\int {d^4 p \over p^2 } ~~ \sim ~~ m^0_f  \int p d p ~~\sim ~~m^0_f 
\Lambda_I^2
\auto\label{trq}
$$
where $m_f^0$ is a bare triplet quark mass. We assume that 
the bare masses that we specifically insert in this way are the only
additional scales apart from $\Lambda_I$. 

\begin{center}

\leavevmode
\epsfxsize=2.5in
\epsffile{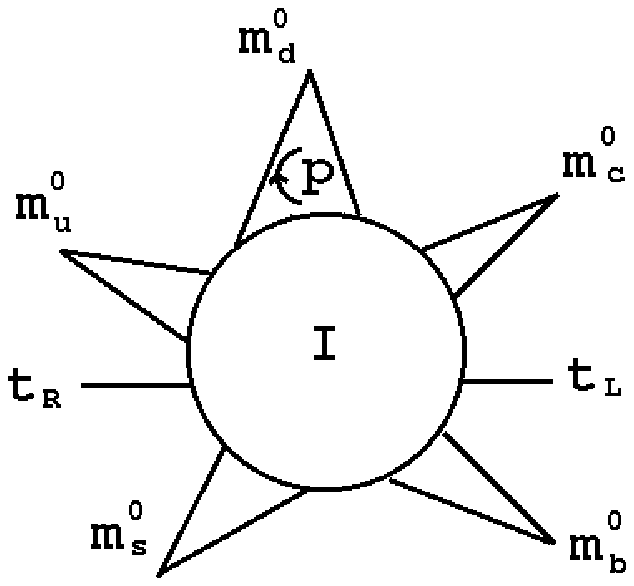}

Fig.~5.3 The Instanton Contribution to the Top Mass. 

\end{center}

The full contribution of interactions of the form of Fig.~5.3 to the triplet 
mass matrix is
$$ 
\eqalign{ \Sigma_3~&\sim ~c_I ~\Lambda_I^{10} ~det~m^0_3 ~[m^0_3]^{-1} \cr
&\sim ~\Lambda_I^{-4} ~det~m^0_3 ~[m^0_3]^{-1} } 
\auto\label{mm} 
$$ 
where $m_3^0$ is specifically the triplet bare mass matrix at the scale 
$\Lambda_I$. After diagonalization (\ref{mm}) is equivalent to
$$
[\Sigma_3]_i ~\sim ~\Lambda_I^{-4} ~\Pi_{j \neq i}~ m_j^0
\auto\label{mm1}
$$
This form is apparent from Fig.~5.3.

A large electroweak scale 
contribution of the form of (\ref{mm}) to the triplet quark mass matrix has
interesting consequences. (\ref{mm}) actually inverts the bare mass 
matrix so that the largest dynamical mass is obtained for that 
quark which has the smallest bare mass. Let us make the (clearly 
oversimplifying) assumption that the bare masses 
are input at the cut-off scale and the single 
instanton interaction represents the dynamical effects of instantons at 
scales between this scale and the electroweak scale. The masses obtained 
from (\ref{mm1}) are then the ``physical'' low-energy masses.
We immediately see that the electroweak scale of the top 
quark mass can be explained as a consequence of it having an 
anomalously small bare mass - potentially a much easier property 
to explain in an extended theory\footnote{The large value of the top quark
mass is well-known to be difficult to obtain in extended technicolor
models\cite{aes}}. Of course, to obtain the complete dynamical evolution of the
high-scale bare masses $m_0^3$ into the dynamical masses given by
(\ref{mm1}) we have to consider much more than the one instanton
interaction we have discussed. Nevertheless if we proceed with the idea that 
the one instanton effect gives a reasonable qualitative picture,
we will have an evolution of the form illustrated in Fig.~5.4. 
\begin{center} 

\leavevmode
\epsfxsize=2.5in
\epsffile{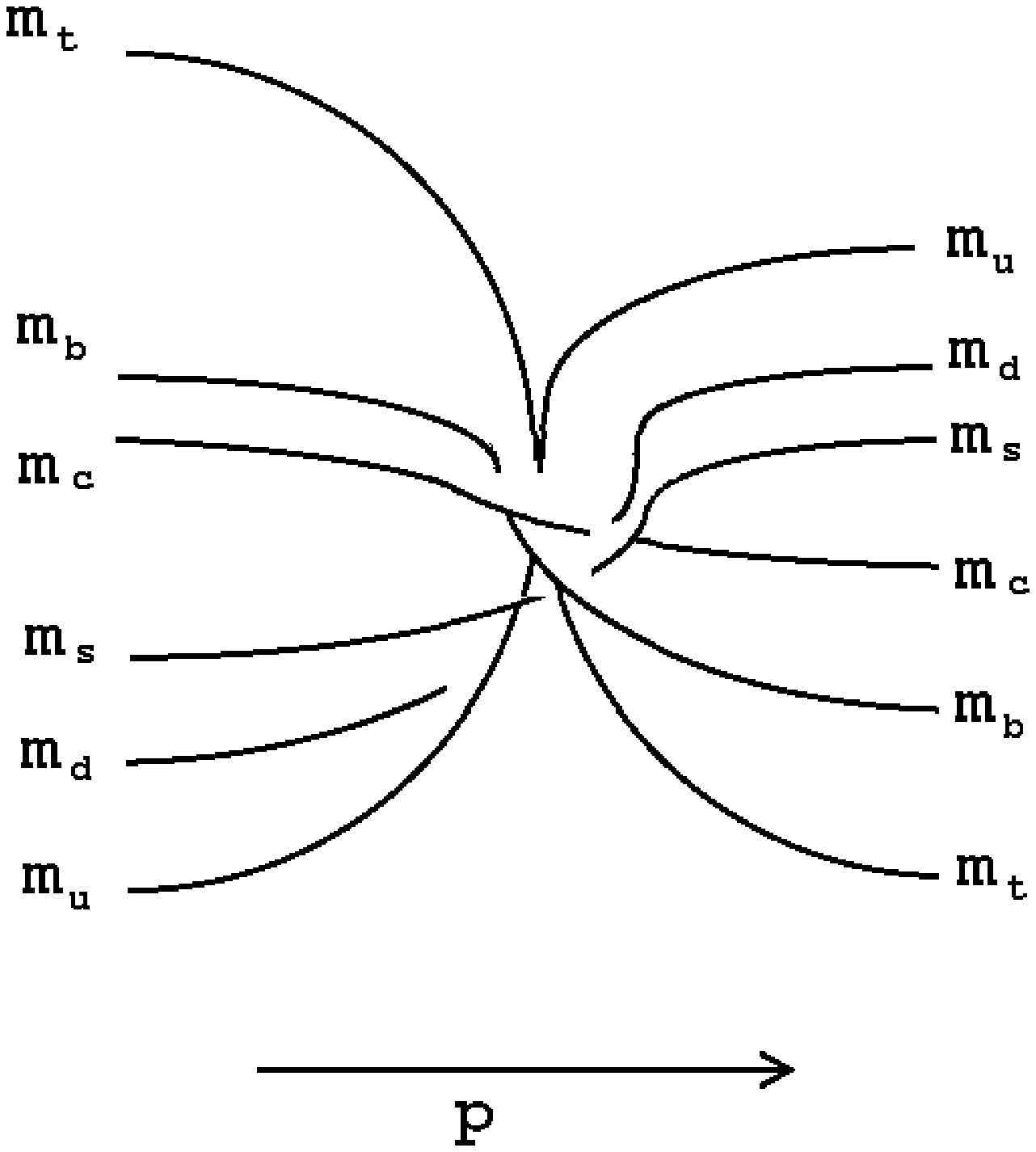}

Fig.~5.4 Inversion of the Mass Spectrum.
\end{center}

In practise it may well be that the complete reversal illustrated in
Fig.~5.4 does not take place. Even so it is clear that the effect of
the instanton interactions is to equalize the masses and so at some point, 
not too far above the electroweak scale, we can expect 
{\it light quark dynamical masses to become 
comparable to that of the top quark mass.} In this case the full
evolution of, say, the $u$ quark dynamical mass, from the triplet chiral
symmetry breaking region up to the electroweak scale and beyond, should be
qualitatively as illustrated in Fig.5.5. 
\begin{center} 

\leavevmode
\epsfxsize=4in
\epsffile{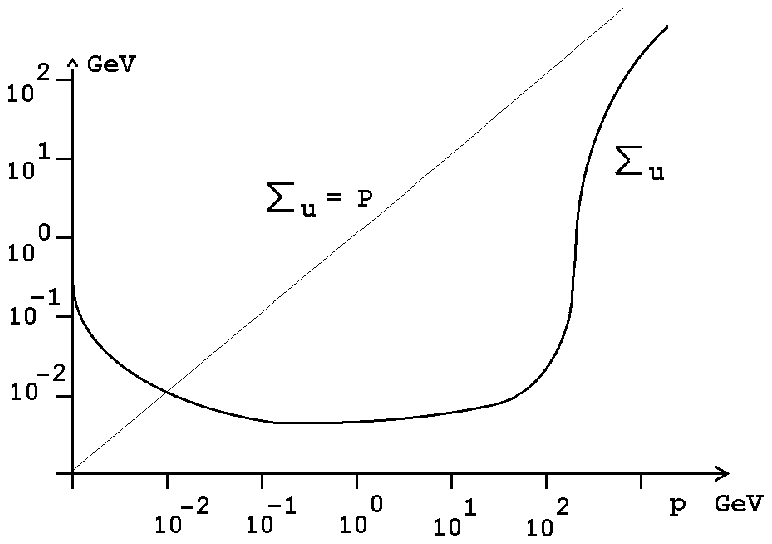}

Fig.~5.5 The $u$-quark Dynamical Mass.
\end{center}
The essence of Fig.~5.5 is that, above the electroweak scale, we 
expect chirality violating amplitudes, in this case the dynamical mass, to 
have the same order of magnitude as the chirality conserving amplitudes, in 
this case the kinematic $\st{p}$ term in the inverse quark propagator. 

In addition to the dynamical masses there will also be four quark
interactions obtained from I involving pairs of quarks with distinct
flavors. These are obtained by closing up all but the quark lines involved 
in the interaction. On the basis that the heavier quarks have smaller bare
masses we expect the largest interaction to be between the heaviest quarks,
i.e. 
$$
V_{bt}~=~\bar{b}_Rb_L\bar{t}_Rt_L
\auto\label{tbI}
$$
involving the top and bottom quark. 
Next in importance will be heavy/light triplet interactions and smallest (at
the electroweak scale) will be the light/light interactions\footnote{These
are the interactions that, at the hadronic scale, contribute to the $\eta '$
mass.}. According to our arguments, all interactions which do not involve
the top will be suppressed by the small top bare mass. 

Consider next sextet/triplet and sextet/sextet four-quark instanton 
interactions. By closing appropriate lines in Fig.~5.1, we obtain 
interactions of the form 
$$
V_{Qq}~=~\bar{Q}_RQ_L\bar{q}_Rq_L ~, ~~~~~~V_{QQ}~=~\bar{Q}_RQ_L\bar{Q}_RQ_L~.
\auto\label{uUI}
$$
If we continue to assume that the one 
instanton interaction represents evolution from the cut-off scale down to 
the electroweak scale then the $V_{Qq}$ interactions are the
electroweak scale ``output'' of the the same higher-scale interactions that
we used to generate bare quark masses. Because of bare top quark mass
suppression, the dominant output interaction will be the sextet/top interaction.
The purely sextet $V_{QQ}$ interaction will also be suppressed by the bare
top quark mass. 

To discuss quark parton model interactions resulting from the instanton 
interactions we consider the ``perturbative'' coupling $G^I_f$ of a
strong or electroweak gauge field $f$ to a scattering triplet quark. 
In Fig.~5.6 we have first separated the coupling into two terms 
$$
G^I_f~=~G^I_{3f}~+~G^I_{6f}  
\auto\label{2co}
$$
according to whether the gauge field couples to a triplet or a sextet
quark. Note that because the 
instanton interaction preserves the sextet vector SU(2) chiral symmetry, 
$G^I_{6f}$ contributes only if $f$ is the singlet hypercharge 
gauge field. As we discuss in the next Section, this will be important if 
(contrary to the arguments we have developed based on the top quark bare 
mass) the $V_{qQ}$ interactions are actually stronger than the $V_{qq}$ 
interactions. We have also shown, in Fig.~5.6, what are the lowest-order
instanton contributions from the four-point interactions if the bare masses
are used for $\Sigma_3$ and the ``perturbative'' dynamical mass discussed in
the last Section is used for $\Sigma_6$. Because of the vector nature of the
elementary gauge coupling there must be a left-right transition involved as 
shown. 

\begin{center}

\leavevmode
\epsfxsize=5in
\epsffile{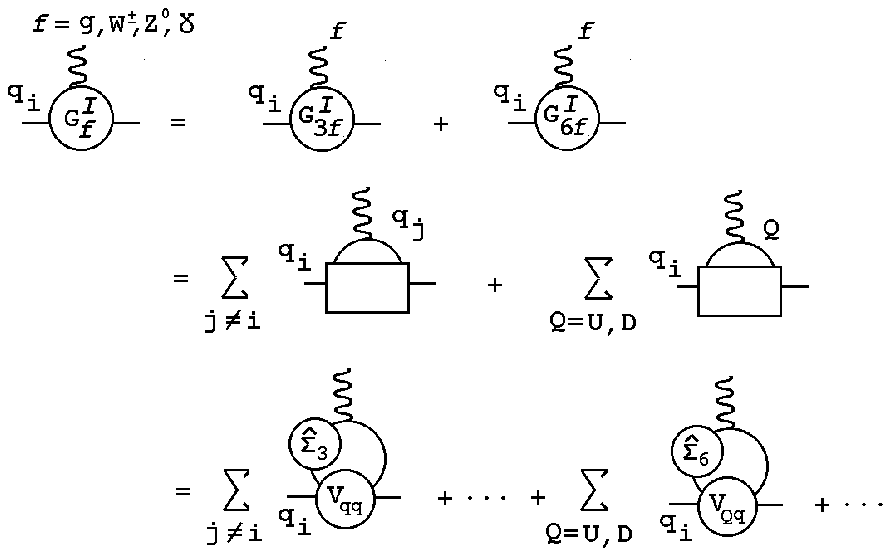}

Fig.~5.6 Gauge Field Coupling.

\end{center}

Since the instanton interactions are (left or right-handed) scalar point 
couplings, the lowest-order $G^I_f$ couplings can be a function of $Q$ only
and so must have the form 
$$ 
G^I_{f,\mu} (P,Q)~=~ G_f(Q^2)~Q_{\mu} 
\auto\label{ico}
$$
In the Landau gauge, for example, this produces no coupling to the gluon or 
photon propagator. For the electroweak bosons, if we write an effective
lagrangian as in Section 2 it is equivalent to the use of unitary gauge, 
in which case it is only the off-shell ``scalar'' components of the $W$ and
$Z$ that we introduce below that couple. Therefore the simple single 
instanton interaction contribution to $G^I_f$ does not provide a 
coupling for the ``physical'' components of the gauge fields.

The full gauge field coupling which is first-order in the instanton 
interaction should also include dynamical mass contributions as in Fig.~5.7.

\begin{center} 

\leavevmode
\epsfxsize=5in
\epsffile{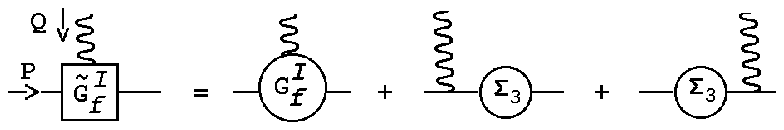}

Fig.~5.7 Inclusion of Dynamical Masses. 

\end{center}
The dynamical mass terms in 
$\tilde{G}_f$ depend only on $P+Q$ and $P$ respectively, giving the
contribution 
$$
G^{\Sigma}_{f, \mu}(P,Q)~=~ \Gamma^f_{\mu} ~{\st{p} +\st{q} \over (p+q)^2 }
~\Sigma_3((p+q)^2) ~+~ 
~\Sigma_3(p^2) ~{\st{p} \over p^2 }~\Gamma^f_{\mu}  
\auto\label{ico1}
$$
where $\Gamma^f$ is the elementary coupling of the gauge field $f$. 

That $G^I_f$ has the simple momentum dependence of (\ref{ico}) is a
consequence of making the one instanton approximation (and of taking only
the 't Hooft point vertex for the instanton interaction.) 
If we consider additional instanton interactions then it is clear 
that $G^I_f$ will have a more general 
momentum dependence and will directly couple to the physical components of 
the gauge field. The simplest ``higher-order'' interactions producing
left-right quark transitions are shown in Fig.~5.8 If the scattering quark
is light then one or more of the internal dynamical mass contributions must
also be that of a light quark (in fact the scattering quark, except when
$f=W^{\pm}$). If this dynamical mass increases as in Fig.~5.5, the most
important contributions 
from the diagrams of Fig.~5.8 will come from the region of loop momenta 
at or above the electroweak scale. Also if both $Q$ and $P$ are at or above the
electroweak scale, then all the momenta entering each of the 
vertices will be at or above this scale and so the diagrams will 
give large contributions. 
\begin{center}

\leavevmode
\epsfxsize=4.5in
\epsffile{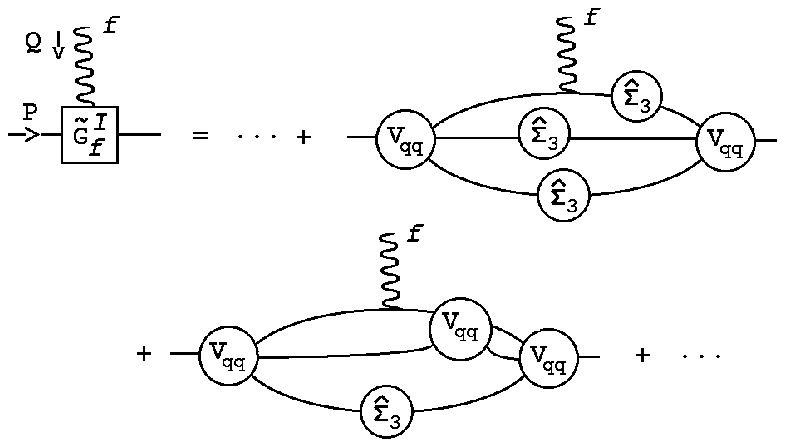}

Fig.~5.8 Higher-order Instanton Couplings.

\end{center}

That the dynamical masses behave as in Figs.~5.4 and 5.5 is 
clearly our starting
point for an understanding of the instanton interactions in general. These
masses give the ``lowest-order'' effects, as in Fig.~5.7, and they are also
essential for ``higher-order'' diagrams of the form of Fig.~5.8 to give
large contributions. It is apparent that replacing
any of $\Sigma_3(p)$ by $\st{p}$ in the diagrams of Fig.~5.8 , will give an
amplitude of comparable magnitude. Consequently it should be a 
general feature of the new interactions introduced by instantons that
chirality changing amplitudes will be comparable in order of magnitude to
chirality conserving amplitudes.

The lowest energy sextet quark instanton interactions involving only the
$W$, $Z$ and $\eta_6$ are inherited directly from sextet pion 
interactions. As we noted in Section 2, the lagrangian derived via the
sextet pion chiral effective theory is equivalent to the standard model 
lagrangian in 
unitary gauge. It is well-known that in unitary gauge it is the
off-shell ``scalar'' components\footnote{These components are often called
``longitudinal''. To avoid confusion, in this paper we will use the word
longitudinal only when referring to physical longitudinal polarizations.}
of the gauge boson fields, i.e. $z^O = \partial^{\mu}Z^0_{\mu},\;
w^+=\partial^{\mu}W^+_{\mu}$ and $w^- = \partial^{\mu}W^-_{\mu}$, that
inherit the interactions of the Goldstone bosons, in this case the
$\pi^0_6,\;\pi^+_6$ and $\pi^-_6$ respectively\cite{gold}. 
In general, there will be instanton interactions coupling $z$'s
and $w$'s to themselves and to the $\eta_6$ of the form $z^0z^0w^+w^-$,
$\eta_6z^0z^0$, $\eta_6w^+w^-$ etc. For example, the $\eta_6w^+w^-$ coupling 
is inherited from the $\eta_6\pi_6^+\pi_6^-$ coupling illustrated in
Fig.~5.9. The derivative nature of all the vector couplings is important.
This implies that they will not contribute to low-energy, on-shell,
interactions of the 
$W$ and $Z$ and so will not give rise to any easily detectable low-energy
deviations from the standard model except, perhaps, if the $\eta_6$, is
involved. Since all the couplings of this kind contain the bare top quark
mass we again expect them to be relatively small. Indeed it appears that the
large value of the top quark mass may actually be responsible for the 
suppression of many of the effects of the sextet sector. 
 
\begin{center} 

\leavevmode
\epsfxsize=4in
\epsffile{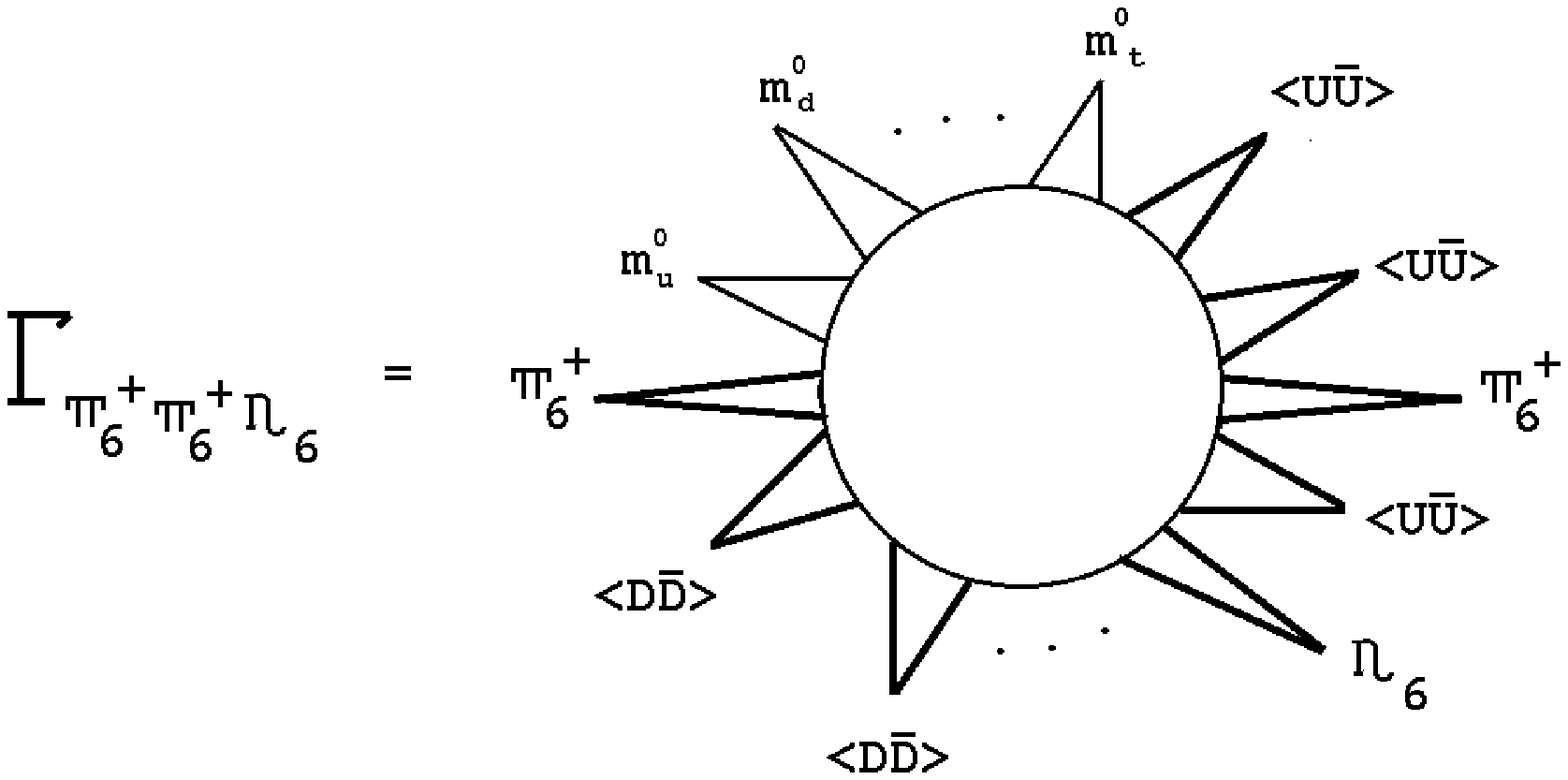}

Fig.~5.9 The $\eta_6\pi_6^+\pi_6^-$ Coupling.
\end{center}

The experimental relevance of the $w$, $z$ and $\eta_6$
vertices is tied to the well-known equivalence theorem\cite{gold,mv}. This 
theorem is commonly applied in the study of strongly-coupled Higgs sector
models. The theorem states that the self-interactions of the Goldstone boson 
sector which appear only in the unphysical 
``scalar'' fields at low energy, appear at high-energy in 
interactions of physical ``longitudinally-polarized'' vector boson 
interactions. In our case this implies that the QCD sextet pion 
couplings will manifest themselves directly in the production and 
interaction of high-energy longitudinally-polarized $W$'s and $Z$'s. 
In effect, at high-energy the scalar $w$'s and $z$'s can be identified with 
physical longitudinally polarized $W$'s and $Z$'s. 

Finally we discuss the $\eta_6$ mass and other properties it should have. 
The bare triplet quark masses or, equivalently, the bare triplet/sextet 
four-quark couplings, break the U(1) symmetry associated with the current
(\ref{cons}) that keeps the $\eta_6$ mass zero. Because of the wide range of 
triplet quark masses at all scales, this symmetry is badly 
broken at all scales. In particular it is badly broken in the momentum range 
where the instanton interactions generate the $\eta_6$ mass and so plays 
essentially no role in keeping this mass small. Since the triplet 
quarks play
only an internal role in the relevant instanton interactions, we can treat the
$\eta_6$ mass as arising from sextet instanton interactions in close
parallel with the usual discussion\cite{gth} of the instanton origin of the
$\eta'$ mass. The one instanton interaction involved is illustrated in
Fig.~5.10. 

To obtain even a crude estimate of the $\eta_6$ mass we must make some 
extremely oversimplifying assumptions. One simplification, which we could 
actually easily 
avoid but which will make the arguments particularly simple, is to
assume that the five light triplet quarks have identical bare masses $m^0$
and physical mass $m$, while the top quark has bare mass $m_t^0$ and
physical mass $m_t$. 
\begin{center}

\leavevmode
\epsfxsize=4in
\epsffile{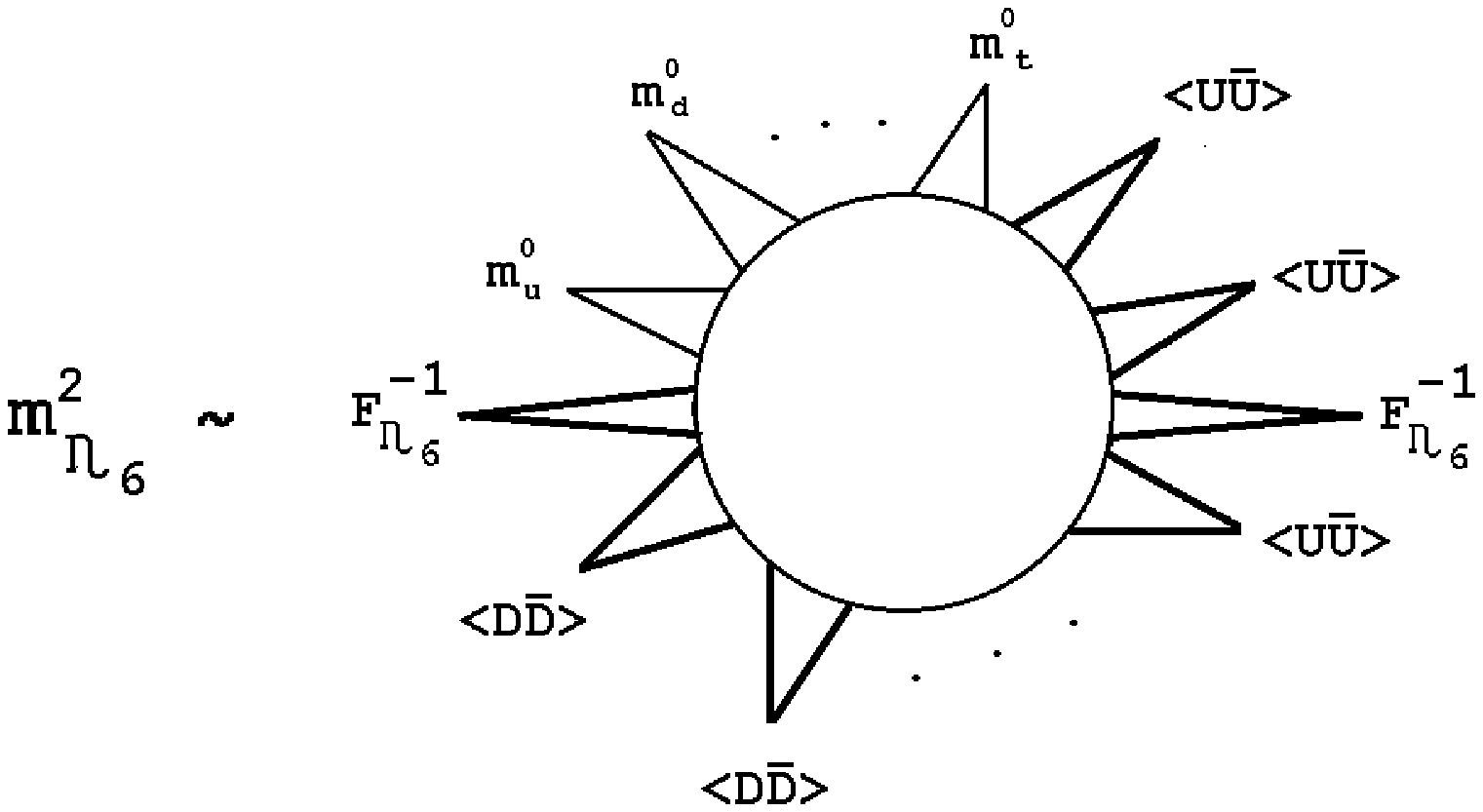}

Fig.~5.10 The Instanton Contribution to the mass of the $\eta_6$.

\end{center}
Assuming that, apart from the explicit mass dependence, the instanton 
interaction has always the same numerical magnitude $N_I$, diagrams of the
form of Fig.~5.3 give
$$
m~\sim~N_I~{m^0_t (m^0)^4 \over \Lambda_I^4} ~, ~~~~~~~~~~
m_t~\sim~N_I~{ (m^0)^5 \over \Lambda_I^4}
\auto\label{ltm}
$$
while Fig.~5.10 gives
$$
m_{\eta_6}^2~\sim~N_I~ {1 \over F_{\eta_6}^2} {m_t^0 (m^0)^5 \over \Lambda_I^2}
\auto\label{nm}
$$
After eliminating $N_I$, we obtain
$$
m_{\eta_6}^2 ~\sim ~{1 \over F_{\eta_6}^2} m~ m^0 ~\Lambda_I^2 
~\sim ~{1 \over F_{\eta_6}^2} m_t~ m_t^0 ~\Lambda_I^2 
\auto\label{nm1}
$$
Taking, say, $F_{\eta_6} \sim 300 $ GeV, $\Lambda_I \sim 10$ TeV,
$m \sim 100$ MeV, $ m^0 \sim 100 $ GeV, (equivalently we could use
$m_t \sim 200 $ GeV, $ m_t^0 \sim 50 $ MeV) we obtain
$$
m_{\eta_6}^2 ~\sim ~ { 10^{-1}~10^2 ~10^{8} \over 10^5} ~\sim 10^4 ~~~GeV^2
\auto\label{nm2}
$$
which gives an electroweak scale mass. Of course, since the $\eta_6$ couples 
to the instanton interactions so directly, we can expect $F_{\eta_6}$ to 
be significantly increased relative to $F_{\pi_6}$. This would 
correspondingly reduce the $\eta_6$ mass. The largest reasonable value 
is probably $F_{\eta_6} \sim 
\Lambda_I$. In this case we obtain a simple estimate, independent of 
$\Lambda_I$, 
$$
m_{\eta_6}~\sim ~ (m~ m^0 )^{- 1 / 2} ~\sim ~ (10)^{- 1 / 2} 
~\sim ~~3~ GeV
\auto\label{nm3}
$$
with the values we have taken for $m$ and $m^0$. Since this number is a 
(crude) lower bound, it seems reasonable 
to expect the $\eta_6$ to have a mass close to, or not too far
below, the electroweak scale. 

It is clear from the above estimates that the raising of the $\eta_6$ mass 
is connected with the large top mass. Even though we have so grossly 
simplified the calculations, it is still apparent that our estimates are
raised by the presence of two mass scales, bare and physical, for
the light quarks (or the top quark). As we have discussed, this is a direct 
consequence of requiring the instanton interaction to give a large top mass.
If we had $m \sim m_0 \sim 10^{-1}$ GeV, then instead of (\ref{nm3}) we would 
obtain $m_{\eta_6} \sim 100 $ MeV, which is considerably lower. 

Above the electroweak scale, the $\eta_6$ will be produced in association with 
the $W$ or the $Z$ via the $ww\eta_6$ and $zz\eta_6$ couplings discussed 
above. It could be seen this way at the Tevatron and would, almost 
certainly, be confused with the standard 
model Higgs. Although we have noted that the $ww\eta_6$ and $zz\eta_6$ 
couplings might be relatively
weak we will not attempt to give any serious estimate of their magnitude. 
The instanton interaction also provides the $\eta_6$ with on-shell couplings 
$\Gamma_{\eta_6,q,\bar{q}}, \cdots~$ 
to quark pairs and higher quark states. The bare mass 
suppression argument implies it should couple most strongly to heavy quark 
pairs. If it is as heavy as we have suggested above, it could decay 
predominantly into  $b\bar{b}$ states. 
(Again this could lead to confusion 
with the standard model Higgs experimently.) It should then be produced in 
$b\bar{b}$ collisions. It will, presumably, also have multi-quark decay
modes. If it is lighter, since it requires an (electroweak scale) instanton 
interaction to be produced or to
decay, it is presumably hard to produce and long-lived once it is produced.

At very high-energy, sextet states will be produced directly and copiously
by QCD multi-gluon (i.e Pomeron) interactions. The higher Casimir implies,
in general, that non-instanton sextet quark interactions are still
``strong'' above the electroweak scale. The production of $W$'s and $Z$'s with
a strong-interaction cross-section will be a striking phenomenon, if seen at
the LHC. 

\mainhead{6.  EXCESS CROSS-SECTIONS}

Consider now the deep-inelastic scattering cross-section at HERA. The kinematics
is illustrated in conventional notation in Fig.~6.1. The primary parton 
model process is a quark $q$ scattering via the exchange of
an electroweak boson $f$. The $f$ can be either a photon, a 
$Z^0$ or a $W^+$. The $\tilde{G}_f$-vertex contains both the standard model
interaction $\tilde{G}^S_f$
and the instanton vertex $\tilde{G}^I_f$. $q(x)$ ($\equiv
~q(x,Q^2)$) is the parton distribution for the quark $q$. 
The excess\cite{HE} cross-section is at very high $x$ 
and $Q^2$, i.e.
$$
x ~\centerunder{\raisebox{1mm}{$\scriptscriptstyle >$}}
{$\scriptscriptstyle \sim$}~0.5 ~~\leftrightarrow ~~xP~\sim 400~GeV,
~~~~~Q~\centerunder{\raisebox{1mm}{$\scriptscriptstyle >$}}
{$\scriptscriptstyle \sim$}~150~GeV
\auto\label{kin}
$$    
\begin{center}

\leavevmode
\epsfxsize=2.5in
\epsffile{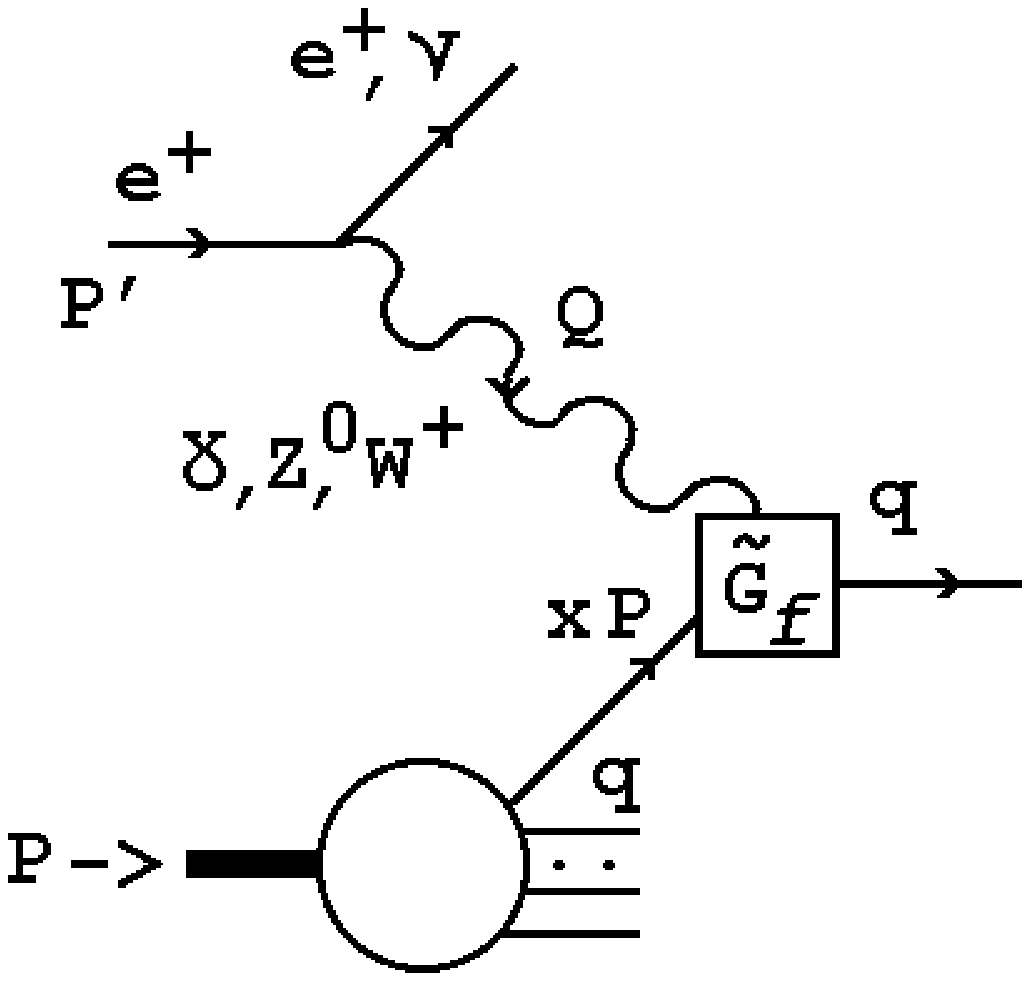}

Fig.~6.1  Deep Inelastic Scattering at HERA.
  
\end{center}
Since both momenta entering the $\tilde{G}_f$
vertex are at the electroweak 
scale, the excess is kinematically just where we expect the instanton vertex 
$\tilde{G}^I_f$ to contribute.

As we discussed in the last Section, the lowest-order $G^I_f$ contribution
gives only vertices for a quark to couple to the $z^0$ and the $w^{\pm}$, 
the ``scalar components'' of the $Z^0$ and $W^{\pm}$. However, the $z^0$
coupling to the positron reduces to 
\beq
\label{small}
\bar{u}(k)\frac{\st{k}}{M^2_{Z^0}}[v_e-a_e\gamma_5]v(k-Q)~
=~-2\frac{m_e}{M^2_{Z^0}}~a_e\bar{u}\gamma_5v
\eeq
where $m_e$ is the positron mass and $u$ and $v$ are positron spinors. 
$v_e$ and $a_e$ are the vector and axial $Z^0$ couplings. The $w^{\pm}$ 
couples in the same manner. Therefore
the  $z^0$ and $w^{\pm}$ exchanges are suppressed by a factor of
$(m_e/M_{Z^0})$ and will not be observed. 

At lowest order in the instanton interaction, therefore, only 
the dynamical mass terms $G^{\Sigma}_f$ contribute to each of the exchanges 
in Fig.~6.1. This gives the two contributions shown in Fig.~6.2. 
In practise we expect the higher-order interactions of Fig.~5.8 etc., to 
give the dominant effect. In particular such processes require both $P$ and 
$Q$ to be at the electroweak scale before they contribute and this is where 
the physical excess cross-section is. The dynamical mass terms are 
functions of $P$ or $P+Q$ only. 
Nevertheless it will be instructive to discuss the 
the dynamical mass contributions. 
Since the processes of Fig.~6.2 are chirality
violating (i.e. helicity-flip) parton processes that are absent in the standard 
model they give cross-sections that should be added to the standard model
parton cross-section. 
\begin{center}

\leavevmode
\epsfxsize=4.5in
\epsffile{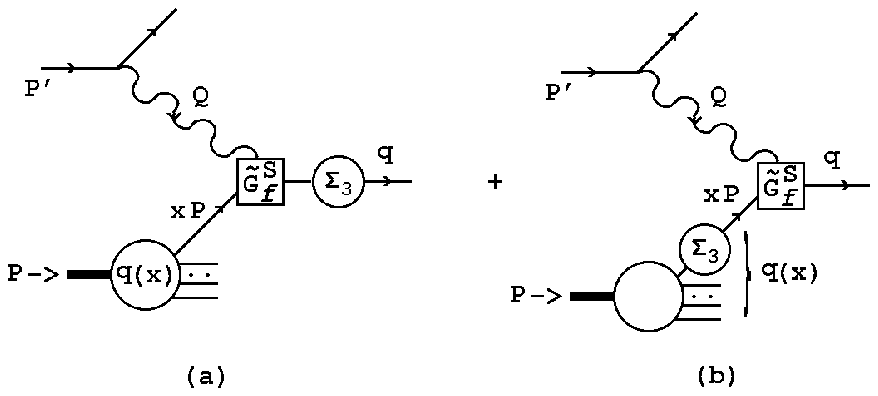}

Fig.~6.2  Dynamical Mass Contributions.
  
\end{center}

We consider first the contribution of Fig.~6.2(a). 
From (\ref{ico1}) we see that we simply obtain
the standard model amplitude $A_{SM}$ multiplied by the quark propagator and
dynamical mass, i.e. 
$$
\eqalign{ A_1^{\Sigma}&~~~~= ~A_{SM}~{x\st{p} +\st{Q} \over (xp+Q)^2 }
~\Sigma_3((xp+Q)^2) \cr
&\centerunder{$\sim$}{\raisebox{-6mm}{$P,Q \to \infty$}}~~
A_{SM} ~{\Sigma_3((xp+Q)^2)\over |xp+Q| }}
\auto\label{dymc}
$$
If we go beyond the lowest-order approximation and iterate and sum 
$\Sigma_3$ interactions (which logically we should not do without considering
other related contributions) then we obtain
$$
A_1^{\Sigma}~~~\centerunder{$\sim$}{\raisebox{-6mm}{$P,Q \to \infty$}}~~
A_{SM} ~{\st{p} + \Sigma_3 \over p^2 - \Sigma_3^2}
\auto\label{dymc1}
$$
This is a trivial example, of course, but we emphasize that
if we are at large enough momentum so that $\Sigma_3 \sim |p|$ then the 
standard model chirality conserving process given by the $\st{p}$
term and the chirality violating process given by $\Sigma_3$ are comparable 
and both are enhanced.

If the final state produces 
a single jet then $(xP +Q)^2$ is the jet mass. 
The experimentally measured jet cross-section
is defined with a fixed-size angular cone and as the transverse energy in 
this cone increases the mass also increases. In the 
conventionally calculated cross-section this is not an important effect.
In contrast, we see from Fig.~5.5 that (\ref{dymc1}) gives a cross-section 
containing an additional factor, compared to the standard model, that will 
increase rapidly with the jet mass as this mass approaches the
electroweak scale. However, since jet masses at HERA are generally less than
10 GeV, this effect will only be significant for the two (or more) jet
cross-section. Even if it is not directly relevant for the HERA excess,
it is nevertheless clear that the dynamical mass contribution of Fig.~6.2(a) 
qualitatively gives an increase in the total cross-section while changing
other kinematic properties very little. 

It is less clear how to systematically discuss the effect of 
Fig.~6.2(b). As we have indicated, the contribution of $\Sigma_3$ should be 
included in the the parton distribution. Logically it should then be 
included generally in the evolution of the distribution. In this case standard
evolution will no longer hold. Since the effect of $\Sigma_3$ will be to
move partons from small momenta out to large momenta and vice versa, it
should spread out distributions in momenta and so enhance large $x$. 
Therefore dynamical mass effects will increase the 
cross-section via the parton distributions in addition to
giving additional final states. Since they represent one instanton effects
in the DIS cross-section we have argued that they
may give a qualitative indication of effects to be expected even if they 
are only a small part of the general electroweak scale 
instanton interactions.

Although multiple instanton interactions are surely essential to
obtain a full picture, and any reasonable order of magnitude 
understanding, of the physics we are discussing, at present we are unable 
to give any sort of quantitative discussion of such effects. 
If we expect chirality violating amplitudes to become
equal in order of magnitude to the existing QCD amplitudes and assume the 
existing amplitudes have their perturbative order of magnitude, this would
crudely say that, if the parton model is still valid, cross-sections should
be four times as large as expected. As in the above discussion of dynamical
mass effects, there is an increase in parton distributions at large $x$, by
a factor of two, and an increase in parton cross-sections, by another factor
of two! Of course, since we have a new set of interactions and a new scale,
we also do not expect the rapid decrease with $Q^2$ embodied in the standard
QCD evolution. 

Consider now the relative order of magnitude of the different gauge bosons 
in fig.~6.2.  Suppose first that the purely
triplet quark interactions $V_{qq}$ are the dominant interactions, as we 
discussed in the previous Section. In this case $G^I_{3f}$ will dominate 
in Fig.~5.6 and in (\ref{2co}).
Since the $\gamma$, $Z^0$ and $W^{\pm}$ 
interactions with the triplet sector are all comparable we would then expect
the neutral and charged currents to have comparable additional 
contributions. 

Alternatively if the sextet interactions are the most important (contrary to 
our arguments based on the smallness of the top quark bare mass) 
a different picture emerges. 
In this case $G^I_{6f}$ will dominate 
in Fig.~5.6 and in (\ref{2co}).
As we noted, only the hypercharge current
has this coupling. From (\ref{lon3}) we also note that the left and 
right-handed components of the hypercharge-current couple to the $\eta_6$.
This implies that the hypercharge 
interaction $G^I_Y$ is uniquely sensitive to the sextet chiral scale.
Correspondingly, we expect the excess to be largest in the hypercharge neutral
current. 

Finally we consider the excess in the inclusive jet cross-section
observed\cite{CDF} at the Tevatron. The remarkable feature of this excess is
that the angular distribution is very close to that of the conventional QCD
cross-section\cite{cdf2}. This in itself suggests that only a ``minor
modification of QCD'' is involved, rather than a major new interaction. The
cross-section 
at large $E_T$ is given mostly by quark-quark scattering and so we discuss 
this first. The angular 
distribution is dominated by the $t$-channel poles due to gluon and quark 
exchange shown in Fig.~6.3.

\begin{center}

\leavevmode
\epsfxsize=4in
\epsffile{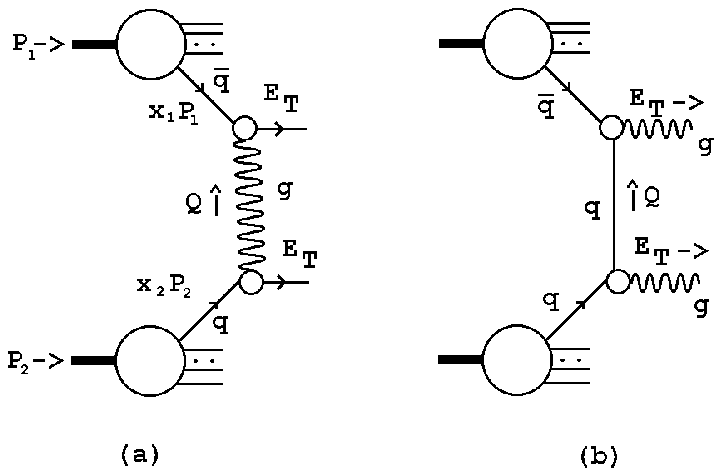}

Fig.~6.3  Quark-Quark Scattering at the Tevatron.
  
\end{center}

A simple explanation of an excess at large $x_1,~x_2$ and 
$Q^2$ would be that $\alpha_s$ is not decreasing as fast as predicted by 
conventional QCD evolution. Indeed in Fig.~4.1(c) we have 
shown just such a behavior for $\alpha_s$ and so 
this would appear to provide a natural explanation of the excess. 
A-priori, however, it would be expected that 
evolution according to the $\beta$-function of Fig.~4.1(c) will only take
place when all momenta are well above the sextet chiral scale. In this case
the lack of $\alpha_s$ evolution is only indirectly the explanation in that
this is responsible for the enhancement of the instanton effects. It could 
also be that we can find a scheme in which we define $\alpha_s$ via jet 
cross-sections and the $\beta$ function of Fig.~4.1(c) can be used down to 
``low energy''. The lack of $\alpha_s$ evolution could then be viewed as the 
essential effect in producing the cross-section excess. 

Amongst the dynamical mass effects in quark-quark scattering 
are the additional scattering processes illustrated in Fig.~6.4 
These contributions will have the same form as (\ref{dymc}), i.e the 
standard model amplitude multiplied by a factor that increases with, one or 
the other of, the jet masses. Such amplitudes trivially preserve the QCD 
angular distribution and also, since jet masses are 
approaching the electroweak scale in the large $E_T$ cross-section, 
give an excess that is increasing 
with $E_T$ relative to the standard model cross-section. 
\begin{center}

\leavevmode
\epsfxsize=4in
\epsffile{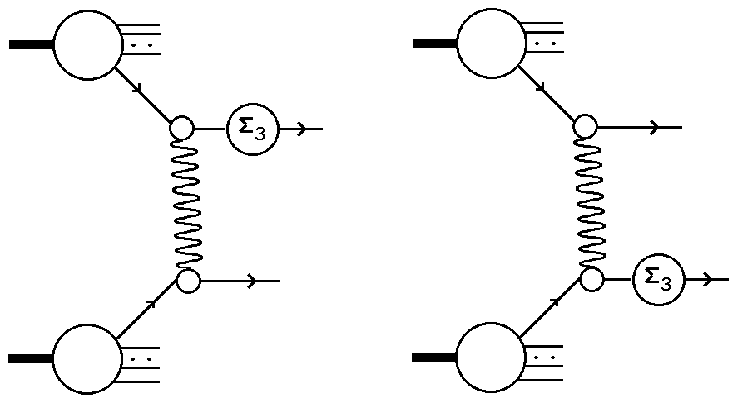}

Fig.~6.4  Dynamical Mass Contributions to Quark-Quark Scattering.
  
\end{center}

There will be no contributions analogous to Fig.~6.4 for the gluon final state
amplitude shown in Fig.~6.3(b). Since this amplitude has a larger color 
factor it follows that the dynamical mass final state effects modify only a 
fraction of the standard model cross-section for quark-quark scattering.
Initial state effects of the dynamical masses will be the same for both 
Fig.~6.4(a) and Fig.~6.4(b). However, if we extend the discussion to 
quark-gluon and gluon-gluon scattering both gluon parton distributions and gluon
final states will not be affected. At the simple dynamical mass level we 
therefore see why the cross-section excess is not as dramatic
as at HERA. DIS scattering at HERA involves only quark partons, while gluons 
are responsible for a significant part of the cross-section at the Tevatron 
and they are not affected.

Again there is very little that we can say about the 
higher-order instanton interactions that we expect to actually give the 
major effect. We can address the simple question of how the cross-section 
can be increased without spoiling the angular distribution. The $t$-channel 
gluon pole of Fig.~6.3(a) is present only in on-shell quark
scattering amplitudes. Off-shell, the pole is cancelled by a zero associated 
with a Ward identity. This pole will, however, be present in the additional
helicity-flip quark amplitudes that we are arguing will be introduced, by
the higher-order vertices of Fig.~5.8 for example. The addition of
helicity-flip on-shell 
quark scattering amplitudes therefore provides an elementary possibility to
increase the cross-section while preserving the angular distribution due to
the $t$-channel gluon pole. In general it is also apparent that the presence of
gluon scattering processes which do not have additional contributions,
will significantly reduce the overall effect as a fraction of the total 
cross-section, compared to HERA. 

\mainhead{7. OTHER RAMIFICATIONS}

If the SQM is the correct description of electroweak symmetry breaking then 
the most obvious prediction is that the full strong interaction will have a 
major strong interaction threshold above the electroweak scale. This should 
be visible in Cosmic Ray physics. Indeed there have been suggestions for 
some time that the ``knee'' in the Cosmic Ray spectrum is actually a strong 
interaction threshold. Since the break is so sharp, this is arguably the 
most rational explanation. It may be coincidence, but the break is in
just the right energy range to be identified with the threshold for the
sextet sector in QCD, as we would like. It is also possible to
argue\cite{arw3} that the variety of exotic effects seen above the knee are
consistent with the SQM. In particular, some of the effects may be
due to the appearance of ``sextet baryons'', containing one sextet quark and
two triplet quarks, with the lowest mass state perhaps being very stable. 

A new high-energy strong interaction sector would surely have major 
implications for early universe physics, particularly if the origin of $CP$ 
violation is in this sector, as we have suggested. If the lowest mass sextet
baryon is neutral it would be a good SIMP (strongly interacting massive 
particle) dark matter candidate. A very stable sextet baryon would also be a 
natural candidate for the extremely high-energy cosmic rays. With a mass of
500 GeV or (probably) higher it would avoid the threshold for
interaction with the cosmic microwave background which rules out\cite{fh}
protons as producing these events. 

The LHC should be well above the sextet threshold and so should see 
ample evidence of the SQM. However, if the SQM is indeed the most immediate
physics beyond the standard model there is little doubt that a higher-energy 
HERA (i.e. e-p collider) would be a better machine to study this physics.
DIS diffractive scattering of the Pomeron with a highly virtual $Z^0$ would 
contain the whole story. 

On the theoretical side, it is not difficult to 
reconcile the SQM with the existing successes of 
both perturbative and lattice $QCD$. The decoupling theorem\cite{ac} assures
us that, at least in short-distance expansions, we can integrate out the 
higher sextet mass scale and apply $QCD$ perturbation theory at current 
momentum scales with only the triplet sector included. Since the infra-red
fixed point value of $\alpha_s$ that we expect to dominate dynamics above
the sextet scale is given by (\ref{as}) there is no problem, at least in
principle, with the idea that integrating out the sextet scale simply
increases $\alpha_s$ from a high-energy value of 1/34 to about 1/8. Clearly
finite size lattice calculations should also remain insensitive to the
higher mass sector. Our belief is that the sextet sector is only relevant if
the full subtleties of the interplay between the infinite volume, chiral and
continuum limits are discussed. 

Study of the QCD Pomeron involves directly reconciling QCD perturbation 
theory with confinement and chiral symmetry breaking. Our study of this 
problem\cite{arw1,arw4} has convinced us that, in general, these properties
are not reconciled\cite{ps}. We believe that the sextet sector, with all its
associated properties, is actually essential for obtaining a consistent
solution to QCD at all energy and transverse momentum scales.

\end{document}